\documentclass[aps,prd,notitlepage,superscriptaddress,nofootinbib,10pt,preprintnumbers]{revtex4-1}
\usepackage{graphicx}
\usepackage{amsmath,amssymb}
\usepackage{hyperref}
\usepackage[usenames]{color}
\usepackage{url}
\usepackage{epstopdf}
\allowdisplaybreaks	

\hypersetup{
    colorlinks=true,
    linkcolor=black,
    citecolor=blue,
    urlcolor=black
}

\def\be{\begin{equation}}
\def\ee{\end{equation}}
\def\ba{\begin{eqnarray}}
\def\ea{\end{eqnarray}}
\def\nn{\nonumber}
\def\f{\frac}
\def\l{\left}
\def\r{\right}

\def\Lzz{\mathcal{L}_{\dot{\zeta}\dot{\zeta}}}
\def\dLzz{{\dot{\mathcal{L}}}_{\dot{\zeta}\dot{\zeta}}}

\begin{document}

\title{A de Sitter limit analysis for dark energy and modified gravity models}

\author{Antonio De Felice}
\thanks{E-mail:antonio.defelice\textit{@}yukawa.kyoto-u.ac.jp}
\affiliation{Center for Gravitational Physics, Yukawa Institute for Theoretical Physics, Kyoto University, 606-8502, Kyoto, Japan}
\author{Noemi Frusciante}\thanks{E-mail:nfrusciante\textit{@}fc.ul.pt}
\smallskip
\affiliation{$^{}$ Instituto de Astrof\text{$\acute{i}$}sica e Ci$\hat{e}$ncias do Espa\c{c}o, Departamento de F$\acute{i}$sica da Faculdade
de Ci$\hat{e}$ncias da Universidade de Lisboa, Edif$\acute{i}$cio C8, Campo Grande, P-1749-016 Lisbon,
Portugal}
\smallskip 
\author{Georgios Papadomanolakis}\thanks{E-mail:papadomanolakis\textit{@}lorentz.leidenuniv.nl}
\affiliation{
$^{}$ Institute Lorentz, Leiden University, PO Box 9506, Leiden 2300 RA, The Netherlands}
\smallskip
\preprint{YITP-17-45}

\begin{abstract}
\vspace{0.2cm}

The effective field theory of dark energy and modified gravity is supposed to well describe, at low energies, the behaviour of the gravity modifications  due to one extra scalar degree of freedom.  The usual curvature perturbation is very useful when studying the conditions for the avoidance of ghost instabilities as well as the positivity of the squared speeds of propagation for both the scalar and tensor modes,   or the St\"uckelberg field performs perfectly when investigating the evolution of linear perturbations. We show that the viable parameters space identified by requiring no-ghost instabilities and  positive squared speeds of propagation does not change by performing a field redefinition, while the requirement of the avoidance of tachyonic instability might instead be different.  Therefore, we find interesting to associate to the general modified gravity theory described in the effective field theory framework, a perturbation field which will inherit the whole properties of the theory. In the present paper we address the following questions: 1) how can we define such a field? and  2) what is the mass of such a field as the background approaches a final de Sitter state? We define a gauge invariant quantity which identifies the density of the dark energy perturbation field valid for any background. We derive the  mass associated to the gauge invariant dark energy field on a de Sitter background, which we retain to be still a good approximation also at very low redshift ($z\simeq 0$).
On this background we also investigate the value of the speed of propagation and we find that there exist classes of theories which admit a non-vanishing speed of propagation, even among the Horndeski model, for which in literature it has previously been found a zero speed.   We finally apply our results to specific well known models. 
 
\end{abstract}
\date{\today}

\maketitle
\flushbottom

\section{Introduction}

The late-time cosmic acceleration questioned our understanding of the gravity force at large scale, thus resulting in a spread of modified gravity theories  and/or dark energy models with the the current aim of going beyond the cosmological standard model, $\Lambda$CDM (we refer the reader to refs.~\cite{Sotiriou:2008rp,Silvestri:2009hh,DeFelice:2010aj,Mukohyama:2010xz,Clifton:2011jh,Tsujikawa:2013fta,Deffayet:2013lga,Joyce:2014kja,Koyama:2015vza,Bull:2015stt} for a complete overview). 

Restricting the attention to those classes of theories which modify the gravitational interaction by including one extra scalar degree of freedom (hereafter DoF), and focusing only on the modifications involving large scale observables, one can handle all the models proposed so far  within the context of the effective field theory of dark energy and modified gravity (EFT)~\cite{Gubitosi:2012hu,Bloomfield:2012ff,Gleyzes:2013ooa,Bloomfield:2013efa,Piazza:2013coa,Frusciante:2013zop,Gleyzes:2014rba,Perenon:2015sla,Kase:2014cwa,Frusciante:2016xoj}, inspired by the EFT of Inflation and quintessence~\cite{Creminelli:2006xe,Cheung:2007st,Weinberg:2008hq,Creminelli:2008wc}. The EFT formalism relies on the unitary gauge for which the additional scalar field, $\phi$, has only a background
profile. An important aspect is the mass of the additional scalar DoF and its impact on the stability of the theory. Recently in literature, the conditions of having an Hamiltonian for the linear perturbations 
bounded from below, have been considered in the context of EFT in the
presence of a dust fluid~\cite{DeFelice:2016ucp}. It was found that it is indeed possible to find
some scalar perturbation variables out of which the Hamiltonian for
the scalar perturbation sector, namely $\textbf{H}(\Phi_i,\dot\Phi_i)$, can be written, in Fourier space, in the following form:
\begin{equation}
\textbf{H}(\Phi_i,\dot\Phi_i)=\frac{a^3}2\left[\dot\Phi_1^2+\dot\Phi_2^2+\mu_1(t,k)\,\Phi_1^2+\mu_2(t,k)\,\Phi_2^2\right],
\end{equation}
where
$\Phi_i(t,k)$ are two linear combinations of the physical fields
$\zeta(t,k)$, the curvature perturbation, and
$\delta\rho_{d}(t,k)$, the perturbation of the dust energy
density. Therefore, a non-negative Hamiltonian would require
$\mu_1$ and $\mu_2$ to be non-negative. This is indeed a required constraint when
we consider the limit for high
$k$, which would correspond to setting positive speed of propagation
for the modes. These constraints would be field-independent. On the
other hand, on looking at larger scales (say $k/(aH)\simeq
1$), the masses of the modes are not any longer negligible and they
do depend on the field one considers. In this
case though, we would like to set proper stability requirements on the value of the mass of physical
perturbation field variables. The above Hamiltonian is written in terms of fields which may not have a clear
physical interpretation. Therefore, in this paper we will try to
address the issue of giving a value for the mass of the perturbation
field which described the energy density of the scalar field $\phi$, i.e. of the dark energy field.

In order to settle the issue regarding the dependence of the mass on the choice of variables we will choose a gauge invariant combination which will describe the perturbations. Then, we will  make a change of coordinates to this new field, $\delta_\phi$,  and proceed to study the mass on the final-state
de Sitter (dS) background. On doing this, we will employ the EFT formalism which allows for late-time dS solutions \textcolor{red}{\cite{Creminelli:2006xe}}. Since we restrict our attention to the dS background, we will
have one, and only one, propagating scalar DoF, because
matter fields are  sub-dominant. Then,  it is possible
to exactly define the speed of propagation and the mass of this
gauge-invariant field representing $\delta_\phi$. Even though the value for the mass of the dark energy field is
exact only on the dS background, it is expected to be a reliable
approximation for its value at late times, i.e.\ when $z\simeq0$, as we live in an universe which is already  dark energy dominated.

Besides the mass we will proceed to investigate, during the dS stage, the
behaviour of the speed of propagation in a model independent fashion. On doing
so we need to consider the limit $k/(aH)\gg1$ as a potential gradient instability might manifest itself at those scales. However, on dS, as time progresses
one needs to consider increasingly  larger values for $k$,  as  $a$ grows exponentially (whereas $H$
remains constant). Subsequently, as the system evolves, the  same modes will be rapidly stretched to cosmological
scales. Now, in general, we find that the speed of propagation for the dark
energy perturbation does not necessarily vanish, even for the Horndeski
subclasses of theories. In fact, the numerical value of the speed of
propagation is model dependent, and its non-negativity can be set as
a constraint in order to have a final stable dS. If this
constraint is not satisfied (i.e.\ $c_s^2<0$) then we will expect that
the late time evolution cannot evolve towards a dS background even
though at the level of the background the dS case is an
attractor solution.  On the other hand, for lower value of $k/(a H)$, the mass of the mode will play a more
important role. In this case one needs to impose, in general, a constraint on the value of the mass for the dark
energy perturbation field in order to obtain a stable dS.

A final source of instability might show up for those theories which exhibit a small or vanishing speed of propagation. In this case the sub-leading order term  in the high $k/(aH)$ expansion becomes relevant and can potentially lead to unstable solutions. We will discuss this in depth and we will present the necessary constraints in order to avoid such instability.

The present paper is organized as follows. In Sec.~\ref{Sec:EFT} we
give a general overview of the EFT approach for dark
energy and modified gravity and we introduce a gauge invariant
quantity to describe the dark energy field. In Sec.~\ref{Sec:parameterspace} we show that the parameter space identified by imposing the no-ghost condition and a positive speed of propagation for scalar modes does not change when considering different quantities describing the dynamics of the extra DoF. In Sec.~\ref{Sec:dS}, we
discuss the dS limit by using the EFT framework, we discuss the
evolution of the extra scalar DoF on different regime, i.e.\ low and
large k, by deriving the speed of propagation and the mass term. In
Sec.~\ref{Sec:examples}, in order to make our results concrete we
apply them to specific well known models, such as K-essence, Galileons and low-energy Ho\v rava gravity. Finally, in
Sec.~\ref{Sec:conclusion} we conclude.

\section{Modifying General Relativity}\label{Sec:EFT}

In the present analysis we will employ a general and unifying approach
to parametrize any deviation from General Relativity obtained by
including one extra scalar DoF in the action,
i.e.\ the effective field theory for Dark Energy and Modified
Gravity~\cite{Gubitosi:2012hu,Bloomfield:2012ff}. For the present
purpose the EFT approach has the advantage of keeping our results very
general and directly applicable to a broad class of theories. Indeed,
all the well known theories of gravity with one extra scalar DoF can
be cast in the EFT framework as shown in refs.~\cite{Gubitosi:2012hu,Bloomfield:2012ff,Gleyzes:2013ooa,Kase:2014cwa,Frusciante:2015maa,Frusciante:2016xoj}.

The EFT is constructed in the unitary gauge, i.e. uniform time hypersurfaces correspond to uniform field hypersurfaces. This results in the scalar perturbation being absorbed by the metric. Let us now introduce the action which can be constructed by solely geometric quantities. The general form is:
\begin{align}\label{EFTaction}
\mathcal{S}^{(2)}&=\int d^4x\sqrt{-g}\l[\frac{m_0^2}{2}(1+\Omega(t))R^{(4)}+\Lambda(t)-c(t)\delta g^{00}+\frac{M^4_2(t)}{2}(\delta g^{00})^2-\frac{\bar{M}^3_1(t)}{2}\delta g^{00}\delta K-\frac{\bar{M}^2_2(t)}{2}(\delta K)^2\nonumber\r.\\
&\l.-\frac{\bar{M}_3^2(t)}{2}\delta K^{\mu}_{\nu}\delta K^{\nu}_{\mu}+\f{\hat{M}^2(t)}{2}\delta g^{00}\delta R^{(3)}+m^2_2(t)\l(g^{\mu\nu}+n^\mu n^\nu\r)\partial_{\mu}g^{00}\partial_{\nu}g^{00}\r],
\end{align}
where as usual $m_0^2$ is the Planck mass, $g_{\mu\nu}$ and $g$ are
respectively the four dimensional metric and its determinant,
$\delta g^{00}=1+g^{00}$, whereas $R^{(4)}$ and $R^{(3)}$ are
respectively the trace of the four dimensional and three dimensional
Ricci scalar, $n_{\mu}$ is the normal vector, $K_{\mu\nu}$ and $K$ are the extrinsic
curvature and its trace. All the operators appearing in the action are
invariant under the time dependent spatial-diffeomorphisms and they
are expanded in perturbations up to second order around a flat
Friedmann-Lema\^\i tre-Robertson-Walker (FLRW) background. The
notation $\delta A= A-A^{(0)}$ indicates the linear perturbation of
the operator $A$ with $A^{(0)}$ its background value.  The
functions appearing in front of each operator are  unknown functions
of time and usually they are named EFT functions. In particular,
$\{\Omega(t), c(t), \Lambda(t)\}$ are called background EFT functions
because these are the only functions that appear in the background Friedmann equations. Finally one can opt to work directly with the field perturbation by restoring the full diffeomorphism invariance,
through the St\"uckelberg technique. This step is useful either when the gauge is not
well defined or when studying the evolution of the perturbations with
numerical tool, such as EFTCAMB/EFTCosmoMC~\cite{Hu:2013twa,Raveri:2014cka,Hu:2014oga,eftweb}.

Now, let us use the Arnowitt-Deser-Misner (ADM)
formalism~\cite{Gourgoulhon:2007ue} and expand the line element around
the flat FLRW background. Keeping only the scalar part of the metric,
we get
 \be\label{scalarMetric} 
ds^2=-(1+2\delta N)dt^2+2\partial_i\psi
dtdx^i+[a^2(1+2\zeta)\delta_{ij}+2\partial_i\partial_j\gamma]\,dx^idx^j
\,, \ee 
where as usual $\delta N(t,x^i)$ is the perturbation of the
lapse function, $\partial_i\psi(t,x^i)$, $\zeta(t,x^i)$ and
$\gamma(t,x^i)$ are the scalar perturbations respectively of $N_i$ and
of the metric tensor of the three dimensional spatial slices,
$h_{ij}$, and $a(t)$ is the scale factor. In the following, since we
choose the unitary gauge, we also set $\gamma(t,x^i)=0$.

It can be shown that the above EFT action can be written as~\cite{DeFelice:2016ucp}
\begin{eqnarray}\label{EFT2}
 \mathcal{S}^{(2)}=\int dt d^3x a^3
&\Biggl\{&
-{\frac {F_4 (\partial^2\psi)^{2}}{2{a}^{4}}}
-\frac32\,F_1\dot\zeta^2+m_0^2(\Omega+1)\,\frac{(\partial\zeta)^2}{a^2}-\frac{\partial^2\psi}{a^2}\left(F_2\delta N-F_1\dot\zeta\right)\nonumber\\
&&{}+4m_2^2\frac{[\partial(\delta N)]^2}{a^2}+\frac{F_3}2\,\delta N^2
+\left[3F_2\dot\zeta-2\l(m_0^2(\Omega+1)+2\hat{M}^2\r)\frac{\partial^2\zeta}{a^2}\right]\delta N\Biggr\},
\end{eqnarray}
where we have defined
\begin{eqnarray}
F_{1} & = & 2m_{0}^{2}(\Omega+1)+3\bar{M}_{2}^{2}+\bar{M}_{3}^{2}\,,\nn\\
F_{2} & = & HF_{1}+m_{0}^{2}\dot{\Omega}+\bar{M}_{1}^{3}\,,\nn\\
F_{3} & = & 4M_{2}^{4}+2c-3H^{2}F_{1}-6m_{0}^{2}H\dot{\Omega}-6H\bar{M}_{1}^{3}\,,\nn\\
F_4 &=& \bar{M}_2^2+\bar{M}_3^2\,, 
\end{eqnarray}
and $H\equiv \dot{a}/a$ is the Hubble function and $\delta N$ and $\psi$ are auxiliary fields. 
 Varying the action with respect to $\delta N$ and $\psi$ yields the Hamiltonian and momentum constraints:
\ba \label{constequations}
&&\frac{2 k^2 \zeta  \left(2 \hat{M}^2+m_0^2 (\Omega +1)\right)}{a^2}+3 F_2 \dot{\zeta}+\frac{8m_2^2 k^2\delta N  }{a^2}+  F_2\frac{k^2 \psi }{a^2}+ F_3\delta N=0\,,\nn\\
&&\delta  N F_2-F_1 \dot{\zeta}-\frac{ F_4}{ a^2}k^2 \psi =0 \,.
\ea 
Finally, solving for the auxiliary fields one can eliminate them from the action, hence obtaining the following Lagrangian, written in compact form  in 3D Fourier space~\cite{Frusciante:2016xoj}:
\be\label{action1}
\mathcal{S}^{(2)}=\int{}d^4x\,a^3\l\{\mathcal{L}_{\dot{\zeta}\dot{\zeta}}(t,k)\dot{\zeta}^2-\f{k^2}{a^2}G(t,k) \zeta^2\r\}\,,
\ee
where 
\be\label{KinGr}
\mathcal{L}_{\dot{\zeta}\dot{\zeta}}(t,k)=\f{\mathcal{A}_1(t)+\f{k^2}{a^2}\mathcal{A}_4(t)}{\mathcal{A}_2(t)+\f{k^2}{a^2}\mathcal{A}_3(t)}\,,\qquad G(t,k)= \f{\mathcal{G}_1(t)+\f{k^2}{a^2}\mathcal{G}_2(t)+\f{k^4}{a^4}\mathcal{G}_3}{(\mathcal{A}_2(t)+\f{k^2}{a^2}\mathcal{A}_3(t))^2}\,,
\ee
are respectively the kinetic and gradient term. The $A_i(t)$ and $G_i(t)$ coefficients are listed in Appendix~\ref{App:list} for a general FLRW background. In the next section they will be specified in the dS limit.

Besides the curvature perturbation $\zeta(t,k)$ one can choose to undo the unitary gauge and work directly with the St\"uckelberg field, namely $\pi$, by performing a broken time translation $t\rightarrow t-\pi(t,\vec{x})$. In order to obtain an unperturbed metric after the translation one needs to recognize that $\zeta = -H \pi$ \cite{Creminelli:2006xe}. However, these fields are not gauge invariant. In this work, we will define a gauge invariant quantity which will describe the evolution of the dark energy field at level of perturbations. Let us introduce the one-form
\begin{equation}\label{normalvector}
n_\mu=\frac{\partial_\mu\phi}{\sqrt{-g^{\alpha\beta}\partial_\alpha\phi\partial_\beta\phi}}=\frac{\delta_\mu^0}{\sqrt{-g^{00}}}\,,
\end{equation}
which would define the 4-velocity along the field-fluid. On the other hand, looking for deviation from General Relativity, when the matter fields are negligible we can can rewrite the Einstein equations as follows
\begin{equation}
m_0^2\,G_{\mu\nu}=T_{\mu\nu}^{\phi}\, .
\label{eq:defTdeS}
\end{equation}
This equation can always be written, and the modifications of gravity have been named in terms of its effective stress-energy tensor, $T_{\mu\nu}^{\phi}$, independently of the EFT which we are considering. Therefore, we can define
\begin{equation}
\rho_\phi \equiv T_{\mu\nu}^{\phi}\,n^\mu n^\nu = m_0^2\,G_{\mu\nu} n^\mu n^\nu\,, \label{eq:defDphi}
\end{equation}
where the second part of this equation holds on-shell, that is, on implementing the equations of motion (at any order). Notice that the definition given in eq.~(\ref{eq:defDphi}) is covariant and, as such, valid even at non-linear order, and does not depend on the choice of the gauge. Since we want the results to match a more phenomenological approach we will define, at linear order the following gauge invariant combination to describe the dark energy field, namely
\begin{eqnarray}\label{gaugeinvariant}
\delta_\phi &\equiv& \frac{\delta\rho_\phi}{{\bar\rho}_\phi}+\frac{{\dot{\bar\rho}}_\phi}{\bar\rho_\phi}
\left[\psi-a^2\,\frac{d}{dt}\left(\frac\gamma{a^2}\right)\right],
\end{eqnarray}
where,  using the background Friedmann equation from action~(\ref{EFTaction}) and assuming that no matter fields are present,  on the background we can define 
\begin{eqnarray}
\bar{\rho}_{\rm \phi}&=&2c-\Lambda-3m_0^2H^2\,(\Omega+a\Omega_{,a})\,,
\label{eq:defrhoDE}
\end{eqnarray}
and 
\begin{equation}
\delta\rho_{\phi}\equiv \rho_{\rm \phi}-{\bar\rho}_{\rm \phi}\, .
\label{eq:defdeltaT00}
\end{equation}
We notice here that $\delta_\phi$ reduces to $\delta\rho_\phi/{\bar\rho}_\phi$ in the Newtonian gauge.  Comma $a$ is the derivative with respect to the scale factor.

We will find the equation of motion for $\delta_\phi$ which in general
assumes the following from
 \ba\label{deltaphieq}
{\ddot{\delta}}_{\phi}+\mu_3(t,k)\,{\dot{\delta}}_{\phi}+\mu_6(t,k)\,{\delta}_{\phi}=0\,.
\ea 
The coefficient of ${\dot{\delta}}_{\phi}$ is the friction term
and its sign will damp or enhance the amplitude of the field
fluctuations. While $\mu_6$ contains both the speed of propagation of
the dark energy field and the information  about of the
mass which, in principle, can be both negative or positive.  The above
equation will allow us to define the mass of the dark energy
perturbation field, which in the next section will be exact on the de
Sitter background, and approximate at low redshifts, $z\simeq0$.

\section{The parameter space for no-Ghost and positive squared speed of propagation for scalar modes}\label{Sec:parameterspace}

By studying the curvature perturbation field, one can immediately work out the stability conditions, namely the no-ghost condition, the positive speed of propagation and the tachyonic condition~\cite{DeFelice:2011bh,Kase:2014cwa,Gleyzes:2015pma,D'Amico:2016ltd,Frusciante:2016xoj,DeFelice:2016ucp}. The first two conditions, i.e.\ the combination of no-ghost and positive-squared-speed conditions, give equivalents constraints for both the $\zeta$ and $\delta_\phi$ fields, in the high-$k$ regime~\cite{Gumrukcuoglu:2016jbh}. We will show it in the following. Let us consider the action~(\ref{action1}) and the field transformation
\be
\delta_\phi=\alpha_3(t,k)\dot{\zeta}+\alpha_6(t,k)\zeta.\label{deltaphidef}
\ee
We will show in the following section that it is possible to derive  this relation and find explicit expressions for $\{\alpha_3,\alpha_6\}$. For the moment we assume that such an expression  exist, since
we have only one independent DoF (the curvature perturbation, $\zeta$), so that
any other field (for example $\delta_\phi$ in this case) can be constructed out of
a linear combination of $\zeta$ and its first time derivative
$\dot\zeta$. Then, on introducing an arbitrary function, $E(t,k)$
(\ note, it is not a field), we can construct the action
 \be
\mathcal{S}^{(2)}=\int{}d^4x\,a^3\l\{\mathcal{L}_{\dot{\zeta}\dot{\zeta}}(t,k)\dot{\zeta}^2-\f{k^2}{a^2}\,G(t,k)
\zeta^2-E(t,k)\,(\delta_\phi-\alpha_3\dot{\zeta}-\alpha_6\zeta)^2\r\},
\label{eq:action2}
\ee
and it is clear that $\delta_\phi$ is a Lagrange multiplier so that we can
use its own equation of motion to remove it from the action. On
performing this step we can see that eq.~(\ref{eq:action2}) reduces to eq.~(\ref{deltaphidef}). This step may look like superfluous, but it allows us to change the
dynamical field variable in the Lagrangian from $\zeta$ to
$\delta_\phi$. Indeed, since $E$ is a free function, if
$\alpha_3\neq0$, on choosing it to be $E=\Lzz/\alpha_3^2$, we
immediately see that the kinetic quadratic term proportional to $\dot\zeta^2$ disappears
and the action can be rewritten, after integrations by parts, as
\begin{eqnarray}
\mathcal{S}^{(2)}&=&\int{}d^4x\,a^3\l\{
 \left[ \frac { \left( H \left( \eta_{\mathcal{L}}-\eta_{{3}}+\eta_{{6}}+3
 \right) \alpha_{{3}}-\alpha_{{6}} \right) \alpha_{{6}}\Lzz}{\alpha_{3}^2}-
\frac{k^2}{a^2}G \right] \zeta^2\right.\nonumber\\
&&{}+\left. \left( -\frac {2\Lzz{\dot\delta}_\phi}{\alpha_{{3}}}+{
\frac { \left( -2\,H \left( \eta_{\mathcal{L}}-\eta_{{3}}+3 \right) \alpha_{{3
}}+2\,\alpha_{{6}} \right) \Lzz\delta_{{\phi}}}{{\alpha_{{3}}}^{2}}
} \right) \zeta-{\frac {{\delta_{{\phi}}^2}\Lzz}{{\alpha_{{3}}^2}}}\r\},
\label{eq:action3}
\end{eqnarray}
where we have defined
\be
\eta_{\mathcal{L}}\equiv\frac{\dLzz}{H\Lzz}\,,\qquad \eta_3\equiv\frac{\dot\alpha_3}{H\alpha_3}\,,\qquad \eta_6\equiv\frac{\dot\alpha_6}{H\alpha_6}\,.
\ee
Therefore, we have succeeded to make $\zeta$ become a Lagrange multiplier and, as such,
in general,
it can be integrated out (using its own equation of motion), leaving $\delta_\phi$ as the propagating independent scalar DoF.

 It should be noted, that integrating out $\zeta$ is only possible whenever the term proportional to $\zeta^2$ in Eq.\ (\ref{eq:action3}) does not vanish. If this case occurs, as we shall see later on happening in some theories for which both $\alpha_6$ and $G$ vanish, then the field $\delta_\phi$ cannot be chosen as the independent field used to describe the system of scalar perturbations.

After removing the auxiliary field $\zeta$, we can rewrite the action as
\begin{equation}\label{actiondeltaphi}
\mathcal{S}^{(2)}=\int d^4x\,a^3 \left[ \frac{a^2}{k^2}\left(Q(t,k)\,{\dot\delta}_\phi^2-\mathcal{G}(t,k)\,\frac{k^2}{a^2}\,\delta_\phi^2\right)\right],
\end{equation}
where the coefficients are listed in Appendix~\ref{App:list}. Therefore, the no-ghost condition for the field $\delta_\phi$ can be read as
\begin{equation}
\lim_{\frac{k}{aH}\to\infty} Q=\lim_{\frac{k}{aH}\to\infty}\frac{\Lzz^2}{G\alpha_3^2}=\frac{\mathcal{A}_3(t)^2}{\mathcal{G}_3(t)}\lim_{\frac{k}{aH}\to\infty}\frac{\Lzz^2}{\alpha_3^2}>0\,,
\end{equation}
which implies 
\begin{equation}
\mathcal{G}_3(t)>0\,,
\end{equation}
and we have assumed that for any function $f(t,k)$ in the Lagrangian,
we have, for large $k$'s, that $f(t,k)=\bar f(t)+\mathcal{O}(k^{-2})$. If
the previous assumption does not hold, then we need to discuss case by
case what happens for the limit. On using again the above assumption,
the speed of propagation can be defined as
\begin{equation}\label{speeddefdelta}
c_s^2=\lim_{\frac{k}{aH}\to\infty} \frac{\mathcal{G}}{Q}=\lim_{\frac{k}{aH}\to\infty}\frac{G}{\Lzz}=\frac{\mathcal{G}_3(t)}{\mathcal{A}_3(t)\mathcal{A}_4(t)}\,,
\end{equation}
which we require to be positive defined. On combining both the constraints we find
\begin{equation}
\mathcal{A}_3(t)\,\mathcal{A}_4(t)>0\,.
\end{equation}

If we consider the stability conditions defined by the field $\zeta$, we find the no-ghost condition
\begin{equation}
\lim_{\frac{k}{aH}\to\infty} \Lzz=\frac{\mathcal{A}_4(t)}{\mathcal{A}_3(t)}>0\,,
\end{equation}
which, together with
\begin{equation}\label{speeddefzeta}
c_s^2=\lim_{\frac{k}{aH}\to\infty}\frac{G}{\Lzz}=\frac{\mathcal{G}_3(t)}{\mathcal{A}_3(t)\mathcal{A}_4(t)}\geq0\,,
\end{equation}
imply $\mathcal{G}_3>0$. Thus, both fields propagate with the same speed.  Note that these results apply on a general FLRW background.

This calculation shows that the no-ghost condition and the speed of
propagation must be calculated in the high-$k$ regime and in such a limit they become
invariants, meaning that they do not change when we change the propagating
scalar DoF. It should be noticed that the no-ghost
conditions do not coincide but the final set of conditions do for 
$\zeta$ and $\delta_\phi$.

Since the mass term is not a quantity which is sensitive to the high
$k$ regime, we should not expect, in general, it behaves as an
invariant. Therefore, each propagating field will have its own mass. However, here we are considering  physical fields, i.e.\ fields
for which we can attach a clear physical meaning and  both
$\delta_\phi$ and $\zeta$ need to remain less than unity for the
background to be stable. Therefore, a mass instability for
$\delta_\phi$, leading this field to reach unity, will imply in
general some instability for the field $\zeta$ and viceversa. 
In order to find the mass of the field $\delta_\phi$ we will investigate
its equation of motion. We will perform this calculation in the
following sections.

\section{The de Sitter Limit}\label{Sec:dS}
 
In this section we will consider the EFT action~(\ref{action1}) in the limit of a dS universe.  Such  a limit is a good approximation in those regimes in which  the dark energy component is dominant over any  matter fluids, e.g.\ very late time.  In this case the background Friedmann equation simply reduces to
\ba\label{eq:dS}
3m_0^2H_0^2=\bar{\rho}_{\phi},
\ea
where the dark energy density, $\bar{\rho}_{\rm \phi}$ has been defined in eq.~(\ref{eq:defrhoDE}).  From the assumption of a dS universe, it follows that $H={\rm const}=H_0$ and the dark energy density is a constant as well. Therefore, eq.~(\ref{eq:defrhoDE}) is a constraint. As a result the dark energy density acts like a cosmological constant. As it is well known such a realization can be obtained, beside the cosmological constant itself, by considering a modified gravity theory with a scalar field whose solution can mimic such a behaviour. Then, eq.~(\ref{eq:dS}) can be integrated and one immediately gets
\be
a(t)=a_0e^{t H_0},
\ee
where $a_0$ is an integration constant.

The EFT approach preserves a direct link with those theories of modified gravity which show one extra scalar DoF and they can be fully mapped in the EFT language~\cite{Gubitosi:2012hu,Bloomfield:2012ff,Gleyzes:2013ooa,Kase:2014cwa,Frusciante:2015maa,Frusciante:2016xoj}. Then, by using the mapping with specific theories and the solution in the dS limit for the chosen theories, we can deduce the behaviour of the EFT functions.  In case of Horndeski~\cite{Horndeski:1974wa} or Generalized Galileon~\cite{Deffayet:2009mn} and beyond Horndeski/GLPV~\cite{Gleyzes:2014dya}, when the shift symmetry is applied, the dS universe can be realized when the kinetic term is a constant, i.e.\ $X=-\dot{\phi}^2=const$~\cite{Nesseris:2010pc,Appleby:2011aa}. In this case all the EFT functions  are constants and the constraint~(\ref{eq:defrhoDE}) is always satisfied. K-essence models~\cite{ArmendarizPicon:2000ah} also admit a dS limit with $\dot{\phi}=const$, when the general function of the kinetic term, namely $\mathcal{K}(X)$, has a polynomial form. In this case the roots of the polynomial obtained by solving the equation $d\mathcal{K}/dX=0$ are the constant values for the derivative of the field.  A more general class of theories is the one with $m_2^2\neq 0$, to which low-energy Ho\v rava gravity~\cite{Horava:2008ih,Horava:2009uw,Mukohyama:2010xz} belongs. Such theory admits a dS solution~\cite{Wang:2010an,Cai:2010hi} and also in this case the EFT functions are constants.  We will assume that the EFT functions on a dS background for all theory having $m_2^2\neq 0$ are constant. In the following, assuming constant EFT functions will greatly simplify the whole treatment. 
  
Moreover, by assuming $\Omega=const$ in the dS limit the EFT background equations reduce to the following forms
\ba
&&3m_0H_0^2(1+\Omega)+\Lambda=0\,, \nn\\
&&3m_0H_0^2(1+\Omega)+\Lambda-2c=0.
\ea
Then, it is easy to deduce the following relations
\ba
c=0\,, \quad \bar{\rho}_\phi=-\f{\Lambda}{1+\Omega}.
\ea

The generality of the EFT approach in describing linear modifications of gravity  due to an extra scalar DoF, allows us to perform a very general analysis in the dS limit for a wide range of theories. However, it is worth to notice that a unique treatment is not possible because subclasses of models, corresponding to specific choices of EFT functions are expected to show up. Therefore, in the following we will mainly consider three subclasses corresponding to 
\begin{itemize}
\item General case: $\{F_4,m^2_2\}\neq0$, to this class belong all models with higher then two spatial derivatives;
\item Beyond Horndeski (or GLPV) models: $\{F_4,m^2_2\}=0$;
\item Ho\v rava gravity-like models:  $m_2^2\neq0$ and $3F^2_2+F_3F_1=0$.   
\end{itemize}   
For all of them we will study the behaviours of the curvature perturbation, $\zeta(t,k)$ as well as of the gauge independent quantity describing the dark energy field $\delta_\phi(t,k)$.

\subsection{The general case}
\label{Sec:generalcase}

We will now investigate the stability of the dS universe in the general case, i.e. by assuming all operators to be active. In contrast to the next cases this corresponds to the case $\{F_4,m^2_2\}\neq0$. The kinetic and gradient terms for this case have the same form as in \eqref{KinGr}, where now the terms $\mathcal{A}_i$ and $\mathcal{G}_i$ are constants and they can be obtained from the time dependent expressions in the Appendix~\ref{App:list} by setting all the EFT functions to be constant. They are:
\ba
\mathcal{L}_{\dot{\zeta}\dot{\zeta}}(t,k)&&=\frac{\left(F_1-3 F_4\right) \left( \left(3 F_2^2+F_1 F_3\right)+8 \f{k^2}{a(t)^2} F_1 m_{2}^2\right)}{2 \left( \left(F_2^2+F_3 F_4\right)+8 \f{k^2}{a(t)^2} F_4 m_{2}^2\right)}\,,\\
G(t,k)&&=\l(16 F_4^2 m_{2}^2 \left(-4 m_0^2 (\Omega +1) \left(m_{2}^2-\hat{M}^2\right)+4 \hat{M}^4+m_0^4 (\Omega +1)^2\right)\f{k^4}{a^4}+8 F_4 \left(4 m_0^2 (\Omega +1) \left(F_2^2 \left(\hat{M}^2-2 m_{2}^2\right)\r.\r.\r.\nn\\
&&\l.\l.\l.+F_3 F_4 \left(\hat{M}^2-2 m_{2}^2\right)+3 \left(F_1-3 F_4\right) F_2 H_0 m_{2}^2\right)+4 \hat{M}^2 \left(F_2^2 \hat{M}^2+F_3 F_4 \hat{M}^2+6 \left(F_1-3 F_4\right) F_2 H_0 m_{2}^2\right)\r.\r.\nn\\
&&\l.\l.+\left(F_2^2+F_3 F_4\right) m_0^4 (\Omega +1)^2\right)\f{k^2}{a^2}+F_2 \left(F_1-3 F_4\right) \left(F_2^2+F_3 F_4\right) H_0 \left(2 \hat{M}+m_0^2 (\Omega +1)\right)\r.\nn\\
&&\l.-\left(F_2^2+F_3 F_4\right){}^2 m_0^2 (\Omega +1)\r)/\l(\left(\left(F_2^2+F_3 F_4\right) +8 F_4 \f{k^2}{a(t)^2} m_{2}^2\right){}^2\r)\,.
\ea

We assume that $\{F_2,\left(F_1-3 F_4\right),2 \hat{M}^2+m_0^2 (\Omega +1)\}\neq 0$, leaving the treatment of these special cases at the end of this section. Now from action~(\ref{action1}), one can derive the field equation for the curvature perturbation, $\zeta$, in the dS limit, which reads 
\be \label{zetaeqgeneral}
\ddot{\zeta}+\l(3H_0+\f{\dot{\mathcal{L}}_{\dot{\zeta}\dot{\zeta}}}{\mathcal{L}_{\dot{\zeta}\dot{\zeta}}}\r)\dot{\zeta}+\f{k^2}{a(t)^2}\f{G}{\mathcal{L}_{\dot{\zeta}\dot{\zeta}}}\zeta=0\,.
\ee

We notice that in the above equation there is no dispersion coefficient. 

Let us now analyse two limiting cases of the above equation. In the limiting case in which $k^2/a^2$ is small, the  term proportional to $\zeta$ in the above equation is sub-dominant and it can be neglected, thus the curvature perturbation behaves as follows 
\be
\zeta (t)= C_2-\frac{C_1 e^{-3 H_0 t}}{3 H_0}\,,
\ee
where $C_i$ are integration constant. Because the second term is a decaying mode, we can deduce from the above result that the curvature perturbation is conserved.

On the contrary, when  $k^2/a^2$ really matters, the equation of $\zeta$ reduces to
\be \label{zetaeqgeneral2}
\ddot{\zeta}+3H_0\dot{\zeta}+(\f{k^2}{a(t)^2}c_{s}^2+\tilde{\mu}_{un})\zeta=0\,,
\ee
where we have defined the squared speed of propagation of the mode $\zeta$ at high-$k$ as in eq.~(\ref{speeddefzeta}) and $\tilde{\mu}_{un}$ is the next to leading order term in the high-$k$ expansion of $G/\mathcal{L}_{\dot{\zeta}\dot{\zeta}}$. We will refer to $\tilde{\mu}_{un}$ as the undamped effective mass of the mode. When considering a dS background these two terms assume the following constant form
\ba
c_{s}^2&&=\frac{\mathcal{G}_3}{\mathcal{A}_3 \mathcal{A}_4}=\frac{F_4 \left(-4 m_0^2 (\Omega +1) \left(m_2^2-\hat{M}^2\right)+4 \hat{M}^4+m_0^4 (\Omega +1)^2\right)}{2 F_1 \left(F_1-3 F_4\right) m_2^2},\\
\tilde{\mu}_{un}&&=-(-12 F_1 \left(F_1-3 F_4\right) F_2 H_0 m_{2}^2 \left(2 \hat{M}^2+m_0^2 (\Omega +1)\right)+F_2^2 \left(3 F_4 \left(-4 m_0^2 (\Omega +1) \left(m_{2}^2-\hat{M}^2\right)+4 \hat{M}^4\r.\r.\\ \nn
&&\l.\l.+m_0^4 (\Omega +1)^2\right)+4 F_1 m_{2}^2 m_0^2 (\Omega +1)\right)+F_1 F_3 F_4 \left(2 \hat{M}^2+m_0^2 (\Omega +1)\right){}^2)/(16 F_1^2 \left(F_1-3 F_4\right) m_{2}^4).
\ea 
Now, let us consider \eqref{zetaeqgeneral2} for a general friction coefficient, $\chi$. Then
for high-$k$, we choose an approximate plane wave solution of the form $\zeta\propto\exp(-i\,\omega\,t)$,
which, after substituting in the previous equation we get the following algebraic equation:
\begin{equation}
-\omega^{2}-\chi iH_{0}\omega+\left(\frac{c_{s}^{2}k^{2}}{a(t)^{2}}+\tilde{\mu}_{un}\right)=0\,,
\end{equation}
The equation has the following solution
\begin{equation}
\omega=-\frac{\chi}{2}\,H_{0}\,i\pm\omega_0\,,
\end{equation}
where
\begin{eqnarray}
\omega_0  \equiv \sqrt{\frac{c_{s}^{2}k^{2}}{a(t)^{2}}+\tilde{m}^{2}}\,, \qquad \tilde{m}^{2}  \equiv \tilde{\mu}_{un}-\frac{\chi^2}{4}\,H_{0}^{2}\,,
\end{eqnarray}
and $\tilde{m}^2$ represents the damped mass of the oscillatory part of the
solution. The imaginary part of $\omega$ corresponds instead to the
decaying (damped) part of the solution. Since we are in the high-$k$
regime, we expect that in general $\frac{c_{s}^{2}k^{2}}{a^{2}}+\tilde{m}^{2}>0$.
In this case we are in the presence of an underdamped oscillator,
for which the solution reads
\[
\zeta(t)\approx e^{-\chi H_{0}t/2}(C_{1}\cos\omega_0 t+C_{2}\sin\omega_0 t)\, ,
\]
and no instability occurs.

Now, an example of where the next to leading order term becomes relevant for stability is when the speed of sound is small or vanishing, i.e.~$c_{s}^{2}\simeq0$. Then, when $\tilde{m}^2<0$,  one has $\frac{c_{s}^{2}k^{2}}{a^{2}}+\tilde{m}^{2}<0$, yielding the following solution:
\begin{equation}
\zeta(t)\approx e^{-\chi H_{0}t/2}(C_{1}e^{-|\omega_0|t}+C_{2}e^{|\omega_0|t})\,,
\end{equation}
which represents overdamped solutions when $|\omega_0|<\chi H_{0}/2$. On the other hand, if 
the model has $\frac{c_{s}^{2}k^{2}}{a^{2}}+\tilde{m}^{2}<0$
and $|\omega_0|>\chi H_{0}/2$, then the mode
\be 
\zeta(t)\propto \,e^{(-\f{\chi}{2}H_0+|\omega_0|)t}
\ee
is exponentially growing. For $c_s^2\simeq 0$ and $\tilde{m}^2<0$ this  implies a catastrophic instability when
\be
 \tilde{\mu}_{un}<0\qquad \text{and} \qquad|\tilde{\mu}_{un}|\gg H_{0}^{2}.
 \ee
 Besides the case described above, when $c_{s}^{2}<0$ and $|\tilde{\mu}_{un}|\simeq H_{0}$ another instability arises. This is then the usual gradient instability. 

The above discussion is directly applicable to the eq. (\ref{zetaeqgeneral2}) presented in this section when $\chi=3$. We will show that the above arguments will be still valid in the high-$k$-limit of the dark energy field for the general case as well as for the other sub-cases discussed in the following, for which one will only need to employ this analysis for different values of $\chi$. In such instances we will refer back to this paragraph instead of repeating the whole discussion.

However, in general the speed of propagation is not vanishing, thus the extra DoF propagates also in a dS universe and the solution, when $\tilde{\mu}_{un}$ is negligible reads:
\ba\label{solzetageneralsubdominant}
\zeta(t,k)=\frac{1}{8 H_0}\l[ \sin \left(\f{k}{a(t)} \f{c_{s}}{H_0} \right)\left(3C_2 H_0+8C_1 \f{k}{a(t)}c_{s}\right)+\cos \left(\f{k}{a(t)} \f{c_{s}}{H_0} \right)\left(8C_1H_0-3C_2 \f{k}{a(t)} c_{s}\right)\right]\,,
\ea
which can be approximated as
\be
\zeta(t,k)\approx \frac{c_{s}}{8 H_0}\f{k}{a(t)}\l[ 8C_1 \sin \left(\f{k}{a(t)} \f{c_{s}}{H_0} \right)-3C_2 \cos \left(\f{k}{a(t)} \f{c_{s}}{H_0} \right)\right].
\ee
This solution decays, even for a very large k as the scale factor  grows exponentially.

Finally, in order to ensure a stable dS universe one has to impose some stability requirements. Following the discussion in the previous section and the results in refs.~\cite{Frusciante:2016xoj,DeFelice:2016ucp}, we have respectively for the avoidance of  scalar and tensor ghosts  
\be\label{generalstab1}
\f{F_1(F_1-3F_4)}{F_4}>0\,, \qquad m_0^2(1+\Omega)-\bar{M}^2_3>0 \,, 
\ee
which need to be combined with the requirement of positive speeds of propagation for scalar and tensor modes
\be\label{generalstab2}
c_{s}^2=\frac{F_4 \left(-4 m_0^2 (\Omega +1) \left(m_2^2-\hat{M}^2\right)+4 \hat{M}^4+m_0^4 (\Omega +1)^2\right)}{8 F_1 \left(F_1-3 F_4\right) m_2^2}\,,\quad c_T^2=1+\f{\bar{M}^2_3}{m_0^2(1+\Omega)-\bar{M}^2_3}.
\ee

At this point one may wonder if a tachyonic condition can be applied. In ref.~\cite{Frusciante:2016xoj}, it has been shown that by performing a field redefinition in order to obtain a canonical action, one can define an effective mass term, which in the small k limit gives the correct condition. If we apply such condition in the dS limit the effective mass associated to our general case is vanishing. Moreover as discussed before, in case the speed of propagation for the $\zeta$ field becomes very small at high-$k$, one has also to ensure that the following conditions do not apply:  $\tilde{\mu}_{un}<0$ and  $|\tilde{\mu}_{un}|\gg H_{0}^{2}$.  However, as already discussed the mass term is sensitive to a field redefinition, thus in order to impose a condition on the mass which holds regardless of the considered field  but containing the real information about the mass of the dark energy field, we need to investigate the behaviour of the gauge invariant quantity $\delta_\phi$.

In the dS universe the gauge invariant quantity defined in eq.~(\ref{gaugeinvariant}) reads
\begin{eqnarray}
\delta_\phi=\frac{\delta\rho_\phi}{{\bar\rho}_\phi} = \frac{2\dot\zeta}H -2\,\delta N -\frac23\,\frac{\nabla^2\zeta+\nabla^2\psi}{a^2 H^2}\,,
\end{eqnarray}
which can be easily obtained from the  first line in eq.~(\ref{constequations}). Moreover, from the same equations we found that $\delta_{\phi}$ can be written as in eq.~(\ref{deltaphidef}) and it is then used to derive the eq.~(\ref{deltaphieq}).  In the dS universe the coefficients of the eqs.~(\ref{deltaphieq})-(\ref{deltaphidef})  are 
\ba\label{coefficientphialphamu}
&&\alpha_3(t,k)=  \f{\tilde{\alpha}_{3}+\f{k^2}{a(t)^2}\f{4}{F_1}\mathcal{A}_4}{3H_0(\mathcal{A}_2+\f{k^2}{a(t)^2}\mathcal{A}_3)}\,,\qquad \alpha_6(t,k)=  \f{2k^2}{3H_0^2a(t)^2}\l[\f{\tilde{\alpha}_{6}}{(\mathcal{A}_2+\f{k^2}{a(t)^2}\mathcal{A}_3)}+1\r]\,,\nn\\
&&\mu_3(t,k)= H_0\f{\sum_{m=0}^{7} b_m \f{k^{2m}}{a^{2m}}}{\sum_{m=0}^7 c_m \f{k^{2m}}{a^{2m}}}\,,\qquad \mu_6(t,k)= \f{\sum_{n=0}^{10} d_n \f{k^{2n}}{a^{2n}}}{\sum_{n=0}^9 f_n \f{k^{2n}}{a^{2n}}}\,,
\ea
where here the $\{b_i,c_i, d_i, f_i, \tilde{\alpha}_i\}$ are constants. Note that the above results might have some limiting cases when the determinants of the above relations go to zero. In what follows we are assuming a non-vanishing denominator.  

For the dark energy field in the regime in which $k^2/a^2$ is negligible, we have 
\be
\mu_3=5H_0+ \mathcal{O}(k^2)\,, \qquad \mu_6=6H_0^2+ \mathcal{O}(k^2)\,,
\ee
where $\mu_6\equiv m^2$ can be read as a mass term, which in this case is positive and of the same order of $H_0^2$, thus no instability takes place. Moreover, because the value of the mass is fixed (i.e. does not depend on the specific value of the EFT functions one can assume), this result is quite general. We also stress that such results can be also safely applicable at low redshifts, as we know at those $z$ the universe is mostly dark energy dominated and thus approaching a dS universe.  Finally, the dark energy field evolves following
\be
\delta_{\phi}(t)=C_1 e^{-3 H_0 t}+C_2 e^{-2 H_0 t},
\ee
and because the friction term is positive, its effect will be to damp the amplitude of the field. Then, in this regime the $\delta_\phi$ field effectively has a mass, while  the $\zeta$ field has not. This is one of the main differences which characterize the gauge invariant field $\delta_\phi$.

 In the opposite regime, we have 
\be \label{GeCahkmu}
\mu_3=7 H_0+\mathcal{O}(k^{-2})\,, \qquad \mu_6= \l(c_{s}^2\f{k^2}{a(t)^2}+\mu_{un}\r) +\mathcal{O}(k^{-2})\,,
\ee
where also in this case we have defined a speed of propagation of the mode $\delta_\phi$ at high-k, which coincides with the speed of propagation for the field $\zeta$ as discussed in Sec.~\ref{Sec:parameterspace} and we have defined, in analogy with the previous case, $\mu_{un}$ as the effective undamped mass for the dark energy filed, which in this case assumes the following form:
\be\label{eqmuundampeddeltaphi}
\mu_{un}=10H_0^2+\f{\mathcal{A}_3(\mathcal{A}_4 \mathcal{G}_2-\mathcal{A}_1 \mathcal{G}_3)+\mathcal{A}_4 \mathcal{G}_3(4\mathcal{A}_4 H_0^2-\mathcal{A}_2)}{\mathcal{A}_3^2 \mathcal{A}_4^2}\,,
\ee
which is the next to leading order term in $\mu_6$. From~\eqref{GeCahkmu} we see that the equation of motion has the form
\begin{equation}\label{deltageneral}
\ddot{\delta}_{\phi}+7H_{0}\dot{\delta}_{\phi}+\left(\frac{c_{s}^{2}k^{2}}{a^{2}}+\mu_{un}\right)\delta_{\phi}=0\,,
\end{equation}
which is exactly the same form of the equation of the $\zeta$ field at high-$k$. Thus the discussion presented earlier is also applicable here, for  $\chi=7$ and $\mu_{un}$ given by eq. (\ref{eqmuundampeddeltaphi}). Finally, the the damped mass of the oscillatory mode is
\begin{eqnarray}
 \hat{m}^{2}  \equiv \mu_{un}-\frac{49}{4}\,H_{0}^{2}\,.
\end{eqnarray}
Therefore, an instability might manifest itself  when $c_{s}^{2}\simeq0$ and  $\hat{m}^2<0$. To be precise,  when one has $\frac{c_{s}^{2}k^{2}}{a^{2}}+\hat{m}^{2}<0$, one must impose $\mu_{un}<0$ and  $|\mu_{un}|\gg H_{0}^{2}$ in order to avoid said instability.

Finally we present the solution at leading order and when $\mu_{un}$ is negligible:
\be\label{deltaphihighkgeneral}
\delta_\phi(t,k)\approx \f{k^3}{a(t)^3} \frac{c_s^3}{1920  H_0^3} \left(1575 c_2 \cos \left(\f{k}{a(t)} \frac{c_s }{ H_0}\right)-128 c_1 \sin \left(\f{k}{a(t)} \frac{c_s }{ H_0}\right)\right)\,,
\ee
which is decaying for an exponentially growing scale factor.

When one considers the case where all the operators are active it is necessary to highlight a number of limiting cases where a different behaviour emerges:
\begin{itemize}
\item case $F_2=0$. In this case, one is still able to solve the constraint equation to write the action in the form~(\ref{action1}), with the following coefficients:
\ba
&&\mathcal{L}_{\dot{\zeta}\dot{\zeta}}=\frac{1}{2} F_1\left(\frac{F_1}{F_4}-3\right)\,,\nn\\
&&G(t,k)=\f{2\f{k^2}{a(t)^2}\l(-4 m_0^2 (\Omega+1) \left(m_2^2-\hat{M}^2\right)+4 \hat{M}^4+m_0^4 (\Omega+1)^2\r)-m_0^2 F_3 (\Omega+1)}{8 \f{k^2}{ a(t)^2} m_2^2+F_3}\,,
\ea
the speed of propagation of the curvature perturbation in the high-$k$ limit ( $k^2/a^2$) is
\be
c_{s}^2=\frac{F_4 \left(-4 m_0^2 (\Omega +1) \left(m_2^2-\hat{M}^2\right)+4 \hat{M}^4+m_0^4 (\Omega+1)^2\right)}{2 F_1 \left(F_1-3 F_4\right) m_2^2}.
\ee
The results and the discussion we had in the general case work also in this case, one has just to replace the correct speed of propagation.
\item case $F_1-3F_4=0$. In this case the kinetic term in action~(\ref{action1}) is vanishing, thus follows that the curvature perturbation $\zeta=0$ as well as the dark energy field. These theories lead to strong coupling thus they cannot be considered in the EFT context. 
\item case $2\hat{M}^2+m_0^2(1+\Omega)=0$. After computing the kinetic and gradient terms, it is straightforward to verify that the gradient term is negative. Indeed, it has the form 
$G=-m_0^2 (\Omega+1)$, and the stability condition to avoid ghost in tensor modes imposes that $1+\Omega>0$. Now, considering that the kinetic terms is positive as well, to guarantee that the scalar modes have no-ghosts, we can conclude that the speed of propagation in negative, thus this subclass of theories in the dS limit shows an instability.
\end{itemize}

In summary, we have analysed the evolution and stability of the curvature perturbation and the gauge invariant dark energy field for a quite general case.
We have found that  the curvature perturbation is conserved at large scale, as expected, and at small scale it  evolves with a non zero speed of propagation, which finally decays as the scale factor grows with time (eq.~(\ref{deltaphihighkgeneral})). The $\delta_\phi$ field at large scale appears to have mass which results to be positive and  of  same order of $H_0^2$, thus avoiding the tachyonic instability and along with the fact that at these scale it decays, these are the two characteristics that makes the two fields analysed to be different.  We conclude this section saying that in order to have a stable dS universe the  conditions which need to be satisfied are the requirements on the kinetic terms  and speeds of propagations for scalar and tensor modes (see eqs.~(\ref{generalstab1})-(\ref{generalstab2})) since the condition on the avoidance of tachyonic instability at large scale is always satisfied. However, one has to make sure that at high-$k$, in case $c_s^2\simeq 0$ the mass associated to these modes do not show an instability, i.e. $\tilde{{m}}^2<0$ when $\tilde{\mu}_{un}<0$ and  $|\tilde{\mu}_{un}|\gg H_{0}^{2}$ for the $\zeta$ field and  $\hat{m}^2<0$ when $\mu_{un}<0$ and  $|\mu_{un}|\gg H_{0}^{2}$ for the dark energy field.

\subsection{Beyond Horndeski class of theories}\label{Sec:beyondH}

In this section we will consider the  EFT action  restricted to the beyond Horndeski class of theories, which corresponds to set $m_2^2=0$, $F_4=0$ in action~(\ref{EFTaction}). 
For such case in general we have both $\mathcal{L}_{\dot{\zeta}\dot{\zeta}}$ and $G$ to be functions of time~\cite{Frusciante:2016xoj}, but in the dS limit the kinetic and the gradient terms  reduce to constants with the following  expressions:
\be
\mathcal{L}_{\dot{\zeta}\dot{\zeta}}= \frac{1}{2} F_1 \left(\frac{F_1 F_3}{F_2^2}+3\right)\,,\qquad G=\frac{F_1 H_0 \left(2 \hat{M}^2+m_0^2 (\Omega +1)\right)-F_2 m_0^2 (\Omega +1)}{F_2}\,,
\ee
and because they are constant we can define the speed of propagation from the beginning without requiring any limit, and it reads
\be\label{speedbh}
c^2_s=\frac{2 F_2 \left(F_1 H_0 \left(2 \hat{M}^2+m_0^2 (\Omega +1)\right)-F_2 m_0^2 (\Omega +1)\right)}{F_1 \left(3 F_2^2+F_1 F_3\right)}\,.
\ee
In the following we will consider $\{F_2, F_1,\left(3 F_2^2+F_1 F_3\right)\neq0\}$. The requirement $F_1\neq0$ is ensured by the assumption that our theory reduces to GR, while the others cases will be considered at the end of this section. The stability conditions requires $\mathcal{L}_{\dot{\zeta}\dot{\zeta}}>0$ and $c^2_s>0$ to guarantee the theory to be free from ghost in the scalar sector and to prevent gradient instabilities. To complete the set of stability conditions one has to include the conditions from the tensor modes~\cite{Frusciante:2016xoj}, i.e. the no-ghost condition which reads  $F_1/2>0$ and a positive tensor speed of propagation, that is $c_T^2=2m_0^2(1+\Omega)/F_1>0$. For the $\zeta$ field we can perform a filed redefinition and construct a canonical action~\cite{Frusciante:2016xoj}, from which we can read the effective mass. In the dS universe, such term is identically zero at all scale.  

In the dS limit the analysis of the dynamical equation for $\zeta$ is straightforward, indeed it is
\be\label{bHzeta}
\ddot{\zeta}+3 H_0 \dot{\zeta}+\frac{k^2}{a(t)^2} c_s^2 \zeta=0\,,
\ee
which has the same form of the equation for $\zeta$ in the general case (see eq.~(\ref{zetaeqgeneral})), thus it has the same solutions in both the regimes, but the speed is now given by eq.~(\ref{speedbh}). In summary, the curvature perturbation is  conserved in the   limit in which $k^2/a(t)^2$ is  heavily suppressed and it slowly decays at high-k (see eq.~(\ref{deltaphihighkgeneral})). 

Now, let us  consider the dark energy field, $\delta_{\phi}$ defined in eq.~(\ref{deltaphidef}).  For the beyond Horndeski sub-case, the coefficients of eqs.~(\ref{deltaphidef})-(\ref{deltaphieq}) reduce as follows 
\ba
&&\alpha_3=-\frac{2 F_1 \left(2 F_2 H_0+F_3\right)}{3 F_2^2 H_0}\equiv \alpha_3^0 \,,\\
&&\alpha_6(t,k)=\frac{2 k^2 \left(-2 F_2 H_0 \left(2 \hat{M}^2+m_0^2 (\Omega +1)\right)+F_2^2\right)}{3 F_2^2 H_0^2 a(t)^2}\equiv \f{k^2}{a(t)^2}\alpha_6^0 \,,\\
&&\mu_3(t,k)=- \frac{H_0 \left(5 \alpha_3^0  \left(\alpha_6^0 H_0-\alpha_3^0 c_s^2\right)-7 (\alpha _6^0)^2 \f{k^2}{a(t)^2}\right)}{\alpha _3^0  \left(\alpha _3^0 c_s^2-\alpha _6^0 H_0\right)+(\alpha _6^0)^2 \f{k^2}{a(t)^2}}\,,\\
&&\mu_6(t,k)=\frac{6 \alpha _3^0 H_0^2  \left(\alpha _3 c_s^2-\alpha _6^0 H_0\right)+\f{k^2}{ a(t)^2} \left[\alpha _6^0 \alpha _3^0 H_0 c_s^2+(\alpha _3^0)^2 (c_s^2)^2+10 (\alpha _6^0)^2 H_0^2\right]+(\alpha _6^0)^2 \f{k^4}{a(t)^4} c_s^2}{\alpha _3^0  \left(\alpha _3^0 c_s^2-\alpha _6^0 H_0\right)+(\alpha _6^0)^2 \f{k^2}{a(t)^2}}\,,
\ea 
where $\alpha_3^0$ and $\alpha_6^0$ are constants. These relations have been obtained from eqs.~(\ref{coefficientphialphamu}), and from them it is easy to identify the $b_i,c_i,d_i$ coefficients. The above expressions hold for $F_2\neq 0$ and $\alpha _3^0  \left(\alpha _3^0 c_s^2-\alpha _6^0 H_0\right)+(\alpha _6^0)^2 \f{k^2}{a^2}\neq 0$. Let us note that in the latter, in order to realize $\alpha _3^0  \left(\alpha _3^0 c_s^2-\alpha _6^0 H_0\right)+(\alpha _6^0)^2 \f{k^2}{a^2}\rightarrow 0$, we have to consider that since all the coefficients are k-independent we need to have $\alpha_6^0=0$, then the remaining option is $c_s^2=0$. That is because  $\alpha_3^0\neq 0$ otherwise the dark energy field disappears. Therefore, the only configuration  is with $\{c_s^2,\alpha_6^0\}=0$. We will consider the case $F_2=c_s=0$ at the end of this section.

In the  limit in which $k^2/a^2$ is suppressed, these coefficients reduce to 
\be
\mu_3=5 H_0+\mathcal{O}(k)\,, \qquad \mu_6=6H_0^2+\mathcal{O}(k^2)\,.
\ee
Then, the friction term $\mu_3$ will dump the amplitude of the dark energy field, while $\mu_6=m^2$ will act as positive dispersive coefficient or a ''mass'' one.  These results are independent on the specific theory one may consider and the mass of the dark energy  field is  positive. This is a general result, which allows us to conclude that all the theories belonging to this sub-class do not experience tachyonic instability in a dS universe, and it is quite safe to assume that this results holds also at $z\approx 0$. Moreover, the solution of eq.~(\ref{deltaphieq}) reads
\be
\delta_{\phi}(t,0)= D_1 e^{-3 H_0 t}+D_2 e^{-2 H_0 t},
\ee
where $D_i$ are integration constant. Therefore, we can conclude that the dark energy field is  damped.

On the other hand, for large $k^2/a^2$ we get
\be 
\mu_3=7 H_0+\mathcal{O}(k^{-2})\,, \qquad \mu_6(t,k)=2 H_0 \left(\frac{\alpha _3^0 c_s^2}{\alpha _6^0}+5 H_0\right)+\frac{k^2 }{a(t)^2}c_s^2+\mathcal{O}(k^{-2})\,,
\ee
with $\alpha _6^0 \neq 0$. Also in this limit the $\mu_3$ coefficient will dump the amplitude of the dark energy field, while the second coefficient assumes the form 
\be
\mu_6(t,k)\equiv \l(\frac{k^2 }{a(t)^2}c_s^2+\mu_{un}\r) \,,
\ee
where the speed of the dark energy field in this regime is the same of the original $\zeta$ field and $\mu_{un}$ follows directly from the previous expression.
The  analysis done  in the previous section for the high-$k$ limit of the dark energy field is directly applicable to this case. Let us just recall that an instability might occurs when at high-$k$ the speed of propagation is very small, as it can happen that $\hat{m}^2<0$ when $\mu_{un}<0$ and  $|\mu_{un}|\gg H_{0}^{2}$. 
 When, $\mu_{un}$ is negligible as in the previous case, we can solve the equation and we find the same behaviour of the general case (eq.~(\ref{deltaphihighkgeneral})).

As before we now separately consider some special cases: 
\begin{itemize}
\item $\l\{c_s^2,\alpha_6\r\}$=0.  In case $c_s^2=0$ the $\zeta$ field has the solution 
\be
\zeta(t)= \tilde{C}_1-\frac{\tilde{C}_2}{3 H_0} e^{-3 H_0 t}\,,
\ee
which predicts the conservation of the curvature perturbation  at any scale. 

When going to the dark energy field, $\delta_{\phi}$, which is related to the $\zeta$ field through the eq.~\eqref{deltaphidef}, one can notice two main aspects. Firstly, because $\alpha_6=0$, the dark energy field  is identified as $\dot{\zeta}$ up to a constant ($\alpha_3$) and hence it requires one boundary condition less. Additionally, when carefully studying the Lagrangian after changing the field, eq.~\eqref{actiondeltaphi}, it is clear that the kinetic term for the dark energy field
diverges for high-$k$. This is due to the fact that the speed is vanishing which translates to the gradient term being zero. Hence, it must be concluded that, for this particular case, the choice for the dark energy field is inappropriate and should not be considered.
\item  $F_2=0$: Considering the action (\ref{EFT2}),  by varying  with respect to $\psi$ immediately follows that  $\dot{\zeta}=0$. Thus the extra scalar DoF does not propagate. 
\item $3F^2_2+F_1F_3=0$: in this case the  kinetic term is zero and the curvature perturbation is vanishing. These theories show strong coupling thus they cannot be considered in the EFT approach.
\end{itemize}

We conclude by saying that the results of the previous section also apply to the beyond Horndeski class of theories considered in the present section. Moreover, the main result here is also that the speed of propagation of the scalar mode in general does not vanish as instead previously found in literature. We will show some practical examples in Sec.~\ref{Sec:examples}. 

\subsection{Ho\v rava gravity like models}\label{Sec:Horavalike}

Let us now consider a special case in which  $m_2^2\neq0$ and $3F^2_2+F_3F_1=0$. In this case the action can be written as
\be\label{specialaction1}
\mathcal{S}^{(2)}=\int{}d^4xa(t)^3\f{k^2}{a(t)^2}\l\{\f{\mathcal{A}_4}{\mathcal{A}_2+\f{k^2}{a(t)^2}\mathcal{A}_3}\dot{\zeta}^2-\l(\f{k^2}{a(t)^2}\f{\mathcal{G}_2+\f{k^2}{a(t)^2}\mathcal{G}_3}{(\mathcal{A}_2+\f{k^2}{a(t)^2}\mathcal{A}_3)^2}+\f{\mathcal{G}_1}{(\mathcal{A}_2+\f{k^2}{a(t)^2}\mathcal{A}_3)^2}\r)\zeta^2\r\}
\ee
with an overall factor $k^2/a(t)^2$.  For this case  in the dS limit the no-ghost and  positive speed conditions read 
\be
\f{\mathcal{A}_4}{\mathcal{A}_3}>0, \qquad c_s^2= \f{\mathcal{G}_3}{\mathcal{A}_3 \mathcal{A}_4}>0\,,
\ee
along with the usual conditions for the stability of tensor modes
\be
 m_0^2(1+\Omega)-\bar{M}^2_3>0 \,,\qquad  c_T^2(t)=1+\f{\bar{M}^2_3}{m_0^2(1+\Omega)-\bar{M}^2_3}>0 \,.
\ee
The conditions on the speeds reduce to $\mathcal{G}_3>0$ and $1+\Omega>0$.
Just for simplicity, let us rewrite the above action as follows 
\be\label{actiontildezeta}
\mathcal{S}^{(2)}=\int{}d^4xa^3\f{k^2}{a(t)^2}\l\{\tilde{\mathcal{L}}_{\dot{\zeta}\dot{\zeta}}(t,k)\dot{\zeta}^2-\l(\f{k^2}{a^2}\tilde{G}(t,k)+\tilde{M}(t,k)\r)\zeta^2\r\}\,,
\ee
where the definitions of the above coefficients immediately follows from the action~(\ref{specialaction1}). 
The  field equation for the curvature perturbation can then be written in a compact form as
\be
\ddot{\zeta}+\l(3H_0+\f{\dot{\tilde{\mathcal{L}}}_{\dot{\zeta}\dot{\zeta}}}{\tilde{\mathcal{L}}_{\dot{\zeta}\dot{\zeta}}}\r)\dot{\zeta}+\l(\f{k^2}{a^2}\f{\tilde{G}}{\tilde{\mathcal{L}}_{\dot{\zeta}\dot{\zeta}}}+ \f{\tilde{M}}{\tilde{\mathcal{L}}_{\dot{\zeta}\dot{\zeta}}}\r)\zeta=0\,,
\ee
where in this case a dispersion coefficient for the field $\zeta$ appears in the evolution equation.  Let us now analyse the two limit as in the previous cases.

In the case $k^2/a^2$ is sub-dominant  $\tilde{M}\neq0$ and we have
\be\label{deltaequationhlike}
\ddot{\zeta}+3H_0\dot{\zeta}+ \bar{m}^2 \zeta=0\,,
\ee
where we have defined the mass term at low k as
\be
\bar{m}^2=\lim_{\f{k^2}{a^2}\to 0 }\f{\tilde{M}}{\tilde{\mathcal{L}}_{\dot{\zeta}\dot{\zeta}}}= \f{\mathcal{G}_1}{\mathcal{A}_2\mathcal{A}_4}\,.
\ee
In order to avoid a instability coming from the mass term we require $|\bar{m}^2|<<H_0^2$.
The solution  reads
\be
\zeta(t)= C_1 e^{\frac{1}{2} t \left(-\sqrt{9 H_0^2-4 \bar{m}^2}-3 H_0\right)}+C_2 e^{\frac{1}{2} t \left(\sqrt{9 H_0^2-4 \bar{m}^2}-3 H_0\right)}.
\ee
When $9 H_0^2-4 \bar{m}^2>0$,  both the exponentials are purely negative hence both modes are decaying. In the opposite case the solution is a decaying oscillator. 

In the limit in which $k^2/a^2$ is dominant the above equation reduces to
\be
\ddot{\zeta}+5H_0\dot{\zeta}+\l(\f{k^2}{a(t)^2}c^2_{s}+ \tilde{\mu}_{un} \r)\zeta=0\,,
\ee
where 
\be
\tilde{\mu}_{un}=\frac{ \left(\mathcal{A}_3 \mathcal{G}_2-\mathcal{A}_2 \mathcal{G}_3\right)}{\mathcal{A}_3^2 \mathcal{A}_4}.
\ee
Let us note that in this limit $\tilde{M}$ is of $\mathcal{O}(k^{-2})$ and the above mass-like term comes from the 0th order expansion of the term $\tilde{G}/\tilde{\mathcal{L}}_{\dot{\zeta}\dot{\zeta}}$.  Also in this case we can apply the analysis of Sec.~\ref{Sec:generalcase}, for $\chi=5$, and conclude that, when $\f{k^2}{a^2}c_s^2+\tilde{m}^2>0$ no instability occurs, while when the speed is small or negligible some growing modes or instability might take place if $\bar{\mu}_{un}<0$. In the case  $\tilde{\mu}_{un}<< \f{k^2}{a^2}c^2_{s}$, the solution of the above equation at leading order is:
\be
\zeta(t,k)\approx -\f{k^2}{a(t)^2} \frac{c_s^2}{96 H_0^2}  \left(45 c_2 \sin \left(\f{k}{a(t)} \frac{c_s }{ H_0}\right)+32 c_1 \cos \left(\f{k}{a(t)} \frac{c_s }{ H_0}\right)\right),
\ee
which decays in time. 

Now, let us consider the dark energy field.
In the regime in which $k^2/a^2$ is sub-dominant, the  equation for the dark energy field has the following coefficients
\be
\mu_{3}=3 H_0+\mathcal{O}(k^{2})\,, \qquad \mu_{6}=\frac{\mathcal{G}_1}{\mathcal{A}_2 \mathcal{A}_4}+\mathcal{O}(k^{2})\,.
\ee
As expected in this case the mass term is the dominant one and $\mu_6\equiv \bar{m}^2$. Thus in this limit the solution is the same of the curvature perturbation.

In the opposite regime, we have
\be
\mu_{3}=9 H_0+\mathcal{O}(k^{-2})\,,\qquad \mu_{6}= \l(\mu_{un}+\f{k^2}{a(t)^2}c_{s}^2\r)+\mathcal{O}(k^{-2})\,,
\ee
where
\be
\mu_{un}=-\frac{-\mathcal{A}_3 \left(6 F_2^2 \mathcal{G}_3 H_0^2+F_3 F_4 \left(6 \mathcal{G}_3 H_0^2+\mathcal{G}_2\right)\right)+\mathcal{A}_2 F_3 F_4 \mathcal{G}_3-14 \mathcal{A}_3^2 \mathcal{A}_4 F_3 F_4 H_0^2}{\mathcal{A}_3^2 \mathcal{A}_4 F_3 F_4}.
\ee
Again here we obtain a behaviour following the one of the $\zeta$ field but with a different dispersive coefficient. When $\mu_{un}$ is negligible the solution at leading order is again an oscillatory decaying mode
\be
\delta_\phi\approx \frac{k^4}{a(t)^4} \f{c_s^4}{53760  H_0^4} \left(99225 C_2 \sin \left(\frac{k}{a(t)} \f{c_s}{H_0}\right)+512 C_1 \cos \left(\frac{k}{a(t)} \f{c_s}{H_0}\right)\right).
\ee

In conclusion, along with the conditions discussed in the beginning of this section for avoiding ghosts and having positive squared speeds of propagations,  we need to make sure that $|\bar{m}^2|<<H_0^2$. Additionally, when the speed of propagation is small, one needs to guarantee that both $\mu_{un}$ and $\tilde{\mu}_{un}$ do not cause an instability. This set of conditions will ensure the system to be stable. We will provide a working example in Sec.~\ref{Sec:examples}, where the above results are applied for low-energy Ho\v rava gravity.

\section{Working examples}\label{Sec:examples}

In this section we will apply the results we have derived in the previous sections to specific models, i.e. K-essence, Horndeski/Galileon models, low-energy Ho\v rava gravity.

\subsection{Galileons}

We consider here the Generalised Galileon Lagrangians, and we will apply the stability conditions derived for the beyond Horndeski models~(Sec.~\ref{Sec:beyondH}). The complete Galileon action is the following~\cite{Deffayet:2009mn}:
\be\label{action:galileon}
\mathcal{S}_{GG}=\int{}d^4x\sqrt{-g}\l(L_2+L_3+L_4+L_5\r),
\ee
where the Lagrangians have the following structure:
\begin{align}\label{GGlagrangians}
L_2&=\mathcal{K}(\phi,X)\,,\nn \\
L_3&=G_3(\phi,X)\Box\phi\,,\nn\\
L_4&=G_4(\phi,X)R-2G_{4X}(\phi,X)\l[\l(\Box \phi\r)^2-\phi^{;\mu\nu}\phi_{;\mu\nu}\r]\,,\nn\\
L_5&=G_5(\phi,X)G_{\mu\nu}\phi^{;\mu\nu}+\frac{1}{3}G_{5X}(\phi,X)\l[\l(\Box\phi\r)^3-3\Box\phi\phi^{;\mu\nu}\phi_{;\mu\nu}+2\phi_{;\mu\nu}\phi^{;\mu\sigma}\phi^{;\nu}_{;\sigma}\r]\,,
\end{align}
here $G_{\mu\nu}$ is the Einstein tensor, $X\equiv\phi^{;\mu}\phi_{;\mu}$ is the kinetic term and $\{\mathcal{K}, G_i\}$ ($i=3,4,5$) are general functions of the scalar field $\phi$ and $X$, and $G_{iX}\equiv \partial G_i/\partial X$.

\subsubsection{The Cubic Galileon model} 

We start by specializing action~(\ref{action:galileon}) to a well known model, i.e. the Cubic Galileon, which corresponds to the following choice of the functions 
\be
K(X)=- \f{g_2}{2} X\,,\quad G_3(X)= \f{g_3}{M^3} X \,,\quad G_4=\f{m_0^2}{2}\,, \quad G_5=0\,,
\ee
where $\{g_2,g_3\}$ are constant and $M^3=m_0H_0^2$ .  

In a dS universe the background equations become
\be
3m_0^2H_0^2= 6 \frac{g_3}{M^3} H_0 \dot{\phi}^3+\frac{1}{2} g_2 \dot{\phi}^2\,, \qquad 3m_0^2H_0^2=-\frac{1}{2} g_2 \dot{\phi}^2\,.
\ee
From the first Friedmann equation one can define the density of the dark energy field at the background, that is 
\ba
\bar{\rho}_{\phi}= 6 \frac{g_3}{M^3} H_0 \dot{\phi}^3+\frac{1}{2} g_2 \dot{\phi}^2\,,
\ea
and  after manipulating the equations, one gets a constraint equation 
\be
6 \frac{g_3}{M^3} H_0 \dot{\phi}^3+ g_2 \dot{\phi}^2=0\,,
\ee
which corresponds to $c=0$, and from which follows~\cite{Deffayet:2009wt,Nesseris:2010pc} 
\be
\dot{\phi}\equiv\dot{\phi}_0=const \,, \qquad g_2=-6\f{g_3}{M^3}H_0\dot{\phi}_0.
\ee
Considering the above results, the EFT functions corresponding to this model in the dS limit read~\cite{Frusciante:2016xoj}
\be
\Lambda=-3 \frac{g_3}{M_3} H_0 \dot{\phi}_0^3=-3m_0^2H_0^2\,,\qquad
M_{2}^4=\f{3}{2}\frac{g_3}{ M^3} H_0 \dot{\phi}_0^3=\f{3}{2}m_0^2H_0^2\,,\qquad
M_1^3=-2 \frac{g_3}{ M^3} \dot{\phi}_0^3=-2m_0^2H_0\,,
\ee
while the others are vanishing. 

Using the mapping and the results obtained in the previous section we obtain that the speed of propagation reduces to zero while the kinetic term diverges, implying that there is no scalar DoF propagating in dS. This is an expected result as the cubic Galileon decouples from gravity in dS with a speed of sound of the form~\cite{Gubitosi:2012hu}
\be
c_s^2=\f{c}{c+M^4_2}\,,
\ee 
which is exactly zero on the background.

\subsubsection{K-essence}

Motivated by the result for the cubic Galileon, where no scalar DoF propagates on dS, we proceed to check if this holds in more Horndeski class theories. According to ref.~\cite{Linder:2015rcz} the complete set of Horndeski models (with $c=0$) does not possess a scalar DoF on a dS background, a statement which we wish to confront with specific examples. 

We start with a well studied and rather simple theory by considering a K-essence model with a general $\mathcal{K}(X)$ and a standard Einstein-Hilbert term.  In this case, we see that
the background equations of motion impose
\be
\mathcal{K}_{,X}|_{X=X_{0}}=0\,,\qquad \mathcal{K}(X_{0})=-3m_{0}^{2}H_0^{2}\,,
\ee
where $X_0$ is the background value of $X$. The speed of propagation can be written, along with the no-ghost condition, as
\begin{equation}
c_{s}^{2}=\frac{\mathcal{K}_{,X}}{2X\mathcal{K}_{,XX}+\mathcal{K}_{,X}},\qquad \mathcal{L}_{\dot{\zeta}\dot{\zeta}}=2X\mathcal{K}_{,XX}+\mathcal{K}_{,X}.
\end{equation}
Now, if $\mathcal{K}$ is analytical, we can consider a Taylor expansion around the point $X=X_{0}$. In such a case  the background equations of motion impose
\begin{equation}
\mathcal{K}=-3m_0^{2}H_0^{2}+\frac{\mathcal{K}_{2}}{2}\,(X-X_{0})^{2}+\frac{\mathcal{K}_{3}}{6}\,(X-X_{0})^{3}+\mathcal{O}[(X-X_{0})^{4}]\,,
\end{equation}
where $\mathcal{K}_{2}\equiv \mathcal{K}_{,XX}(X_{0})$, and $\mathcal{K}_{3}\equiv \mathcal{K}_{,XXX}(X_{0})$.
Then, one finds that
\begin{equation}
c_{s}^{2}=\frac{1}{2X_{0}}\,(X-X_{0})-\frac{1}{4\mathcal{K}_{2}X_{0}^{2}}\,(3\mathcal{K}_{2}+X_{0}\,\mathcal{K}_{3})\,(X-X_{0})^{2}+\mathcal{O}[(X-X_{0})^{3}]\,,
\end{equation}
and  we have that, for an analytical function, $c_{s}^{2}\to0$ on dS. Hence, if one would want to design a K-essence model with a non zero speed of sound one  has to resort to a non analytic form for $\mathcal{K}$. Therefore, this is an example for which in the class of Horndeski models it is still possible to have a propagating DoF in the dS universe. In the following we will show more.

\subsubsection{Covariant Galileons}

Let us study the dS solution for the Covariant Galileon~\cite{Deffayet:2009wt}, defined by the following choice of the functions:
\be
K(X)  =  \frac{c_{2}}{2}\,X\,, \quad G_{3}(X)  =  \frac{c_{3}\,X}{2M^{3}}\,,\quad G_{4}(X) = \frac{m_{0}^{2}}{2}-\frac{c_{4}X^{2}}{4M^{6}}\,,\quad G_{5}(X) =  \frac{3c_{5}\,X^{2}}{4M^{9}}\,,
\ee 
We proceed by adopting the following definitions~\cite{DeFelice:2010pv} 
\be
X  = -x_{dS}^{2}\,m_{0}^{2}\,H_0^{2}\,,\quad  \alpha  \equiv  c_{4}\,x_{dS}^{4}\,,\quad \beta  \equiv  c_{5}\,x_{dS}^{5}\,,
\ee
where $x_{dS}=\frac{\dot{\phi}_0}{m_0H_0}|_{dS}$ being the dS solution and M has been defined before. Then, we find that the equations of motion for the background are fulfilled
provided that
\begin{equation}
c_{2}\,x_{dS}^{2}=9\alpha-12\beta+6\,,\qquad c_{3}x_{dS}^{3}=9\alpha-9\beta+2\,.
\end{equation}

In this case  the no-ghost condition for the scalar mode
can be written as
\begin{equation}
\frac{\mathcal{L}_{\dot{\zeta}\dot{\zeta}}}{m_0^{2}}=-\frac{\left(3\,\alpha-6\,\beta+2\right)\left(3\,\alpha-6\,\beta-2\right)}{6\left(\alpha-2\,\beta\right)^{2}}>0\,,
\end{equation}
and the speed of propagation reduces to
\begin{equation}
c_{s}^{2}=\frac{\left(2\,\beta-\alpha\right)\left(15\,\alpha^{2}-48\,\alpha\,\beta+36\,\beta^{2}+4\right)}{18\,\alpha^{2}-72\,\alpha\,\beta+72\,\beta^{2}-8}\,,
\end{equation}
which does not vanish in general. Finally, from the dark energy field sector, we obtain
\be
\mu_{un}=-\frac{4 H_0^2 \left(15 \alpha ^3-6 \alpha ^2 (13 \beta +5)+2 \alpha  \left(66 \beta ^2+57 \beta +17\right)-4 \left(18 \beta ^3+27 \beta ^2+17 \beta +3\right)\right)}{(-3 \alpha +6 \beta +2)^2}\,,
\ee
which must be constrained, as  discussed before,  in the case of a vanishing speed of propagation.
 Correspondingly, we obtain for the tensor sector the following:
\begin{equation}
\frac{A_{T}^{2}}{m_0^{2}}=\frac{1}{8}\left(3\,\alpha-6\,\beta+2\right)>0,\qquad
c_{T}^{2}=\frac{\alpha-2}{6\,\beta-3\,\alpha-2}\,.
\end{equation}
Considering the no-ghost condition and a positive speed of propagation  it can be easily shown that a part  of the parameter space allows for stable dS solutions with a non-vanishing speed of propagation. For example the choice\ $\alpha=-\tfrac{7}{5}$, and $\beta=-\tfrac{4}{5}$ achieves this. These values result in a relatively small speed of propagation for which $\mu_{un}>0$. Thus no instability is present for these choice of parameters.

\subsubsection{Models with $G_{5}(X)=0$, and $G_{4}(X)=m_0^2/2$}

Now, for the Covariant Galileon,  setting $\alpha=0=\beta$, that is $G_{4}=m_0^{2}/2$
and $G_{5}=0$, yields once more a vanishing $c_{s}^{2}$    while for the kinetic term implies $\mathcal{L}_{\dot{\zeta}\dot{\zeta}}\to{+}\infty$,
i.e.\ weak coupling regime (see the Cubic Galileon case in the previous section). Therefore, the Covariant Galileon requires
non-trivial $G_{4},G_{5}$ in order to have a non-zero speed of propagation
for the scalar modes.

It is possible to find models for which $G_{4}=m_0^{2}/2$
and $G_{5}=0$, and, on dS, the speed of propagation does not
vanish. We illustrate this by  considering the model:
\be
K(X) =  -c_{2}\mu^{4}\left(\frac{-X}{2M^{4}}\right)^{p}\,,\quad G_{3}(X)  =  c_{3}\,\mu\left(\frac{-X}{2M^{4}}\right)^{q},\quad G_{4}(X)  =  \frac{m_0^{2}}{2}\,,\quad
G_{5}(X)  = 0\,,
\ee
where $p$ and $q$ are constants and $\mu$ is a typical length scale of the system.
Using the same notation as for the Covariant Galileon we obtain from  the background equations of motion the following:
\begin{equation}
c_{2}=\frac{3m_0^{2}H^{2}}{\mu^{4}(-X/(2M^{4}))^{p}}\,,\qquad c_{3}=-\frac{p\,m_0^{2}\,H}{\,\mu\,q\,(-X)^{1/2}\,(-X/(2M^{4}))^{q}}\,.
\end{equation}
and subsequently we obtain:
\begin{eqnarray}
\frac{\mathcal{L}_{\dot{\zeta}\dot{\zeta}}}{m_0^{2}}  =  \frac{3p\,(1-p+2q)}{(1-p)^{2}}\,,\quad c_{s}^{2}  =  \frac{1-p}{3(1-p)+6q}\,,\quad \mu_{un}=\frac{2 H_0^2 \left(21 p^2-2 p (18 q+11)+1\right)}{3 p (p-2 q-1)}\,,
\end{eqnarray}
whereas the tensor modes do not add any new constraint. It is possible
to find a stable dS on choosing $0<p<1$, and $q>-\tfrac{1}{2}\,(1-p)$. Finally, in order for the $\mu_{un}$ term to create an instability, one needs to look at the case of a very small (or vanishing) speed of sound, i.e. $p\rightarrow 1$ or $q\rightarrow \infty$. In both cases it turns out that $\mu_{un}=12H_0^2$ hence no issues arise.  As an example for the choice of parameters, we choose $p=1/2$ and $q=2$, for which all the conditions are satisfied with a speed of propagation of $c_s^2=1/27$, and the undamped mass of the modes is not negligible, as $\mu_{un}=326/27\,H_0^2$.

Therefore we have showed that, even in the absence of non-trivial
$G_{4},G_{5}$, it is still possible to find models for which $c_{s}^{2}$
does not vanish on dS. This concludes our demonstration of the fact that  Horndeski models do not necessarily imply a vanishing DoF on a dS background as suggested by ref.~\cite{Linder:2015rcz}.

\subsection{Low-energy Ho\v rava gravity}\label{Horava}

One well known model which falls in the above sub-case is the low-energy  Ho\v rava gravity~\cite{Horava:2008ih,Horava:2009uw,Mukohyama:2010xz}. The action of this theory is
\begin{eqnarray} \label{horavaaction}
\mathcal{S}_{H}=\f{1}{16\pi G_H}\int{}d^4x\sqrt{-g}\left(K_{ij}K^{ij}-\lambda K^2 -2 \xi\bar{\Lambda} +\xi \mathcal{R}+\eta a_i a^i\r),
\end{eqnarray}
$\left\{\lambda,\xi,\eta\right\}$ are dimensionless running coupling constants, $\bar{\Lambda}$ is the ``bare'' cosmological constant, $G_H$ is  the coupling constant which can be expressed as~\cite{Blas:2009qj}
\be
\f{1}{16\pi G_H}=\f{m_0^2}{(2\xi-\eta)}.
\ee

Expanding the above action in terms of the perturbed metric~(\ref{scalarMetric}) and considering the mapping between this action and the EFT framework, the action up to second order in perturbations can be recast in the the same form of action~(\ref{specialaction1}) and by using the redefinition~(\ref{deltaphidef}) the action becomes the one in~(\ref{actiontildezeta}). In order to specify the coefficients for action~(\ref{actiontildezeta}) and then analyse the solutions for this specific model, let us consider the background equation which in the dS limit is
\be
H_0^2=\f{2\xi\bar{\Lambda}}{3(3\lambda-1)}\,,
\ee
from which follows
\be
\bar{\rho}_{\phi}=m_0^2\f{2\xi\bar{\Lambda}}{(3\lambda-1)}.
\ee
Now, we can specify all the EFT functions~\cite{Frusciante:2015maa},
\begin{align}\label{Horava_mapping}
&(1+\Omega)=\f{2 \xi}{(2\xi-\eta)}, \qquad \Lambda=-\f{4m_0^2\xi^2\bar{\Lambda}}{(2\xi-\eta)(3\lambda-1)}\qquad\bar{M}_3^2= -\f{2m_0^2}{(2\xi-\eta)}(1-\xi),\nn \\
&\bar{M}_2^2 =-2\f{m_0^2}{(2\xi-\eta)}(\xi-\lambda), \qquad m^2_2=\f{m_0^2 \eta}{4(2\xi-\eta)},\qquad\bar{M}_1^3=\hat{M}^2=c=M_2^4=0.
\end{align}
Then, the no-ghost and gradient conditions at high-$k$ read 
\be
\frac{2 (1-3\lambda)}{(\lambda -1) (\eta -2 \xi )}>0,\qquad c_{s}^2=\frac{(\lambda -1) \xi  (2 \xi-\eta )}{\eta  (3 \lambda -1)}>0\,,
\ee
where the latter is different from zero even in the PPN limit ($\eta\rightarrow 2\xi-2$). Additionally,  when $k/a$ is sub-dominant we obtain a vanishing  mass term for the $\zeta$ field, i.e. $\bar{m}^2=0$.

When $k/a$ is dominant, we also need to consider the undamped mass for the $\zeta$ field, which is 
\be\label{mtun}
\tilde{\mu}_{un}=\frac{4 H_0^2 \xi}{\eta }.
\ee 
When studying the parameter space allowed further below it turns out that $\tilde{\mu}_{un}$ will remain manifestly positive, hence no instabilities will occur due to its presence.
Now, when it comes to the gauge independent choice, the dark energy field, $\delta_{\phi}$ adds no new conditions when demanding  no-ghost and a positive speed of propagation as analysed in the previous section. The mass for this field at low k is vanishing as well. 
At high-$k$ for the dark energy field we can define
\be \label{mun}
\mu_{un}=\frac{2 H_0^2 (\eta  (21 \lambda +2 \xi -7)+2 \xi  (3 \lambda -2 \xi -1))}{\eta  (3 \lambda -1)}\,,
\ee
which has to be constrained if the speed is very small. Further below we will comment on its effect on the parameter space.
Finally the tensor sector add the following set of constraints to the model:
\be 
\frac{2 }{2 \xi-\eta }>0, \qquad c_T^2=\xi>0.
\ee
Now it is possible to define a range of viability for the parameters of low-energy Ho\v rava gravity based on this set of conditions, namely:
\be
0<\eta<2\xi,\qquad \lambda>1\quad \text{or}\quad \lambda<\f{1}{3}.
\ee
Keeping the above conditions in mind we turn our attention to the regime of a small speed, i.e. $\lambda\rightarrow 1$ or $\eta\rightarrow 2\xi$. In both cases it is easy to see that \eqref{mtun} will always be positive. On the contrary, \eqref{mun} does show different behaviours. For $\lambda\rightarrow 1$ it is clear that an increasing $\xi$ pushes it more and more to the strongly negative regime while $\eta$ does the opposite. Now, for $\eta\rightarrow 2\xi $, it reduces to a constant, $\mu_{un}=16H_0^2$.

\section{Conclusion}\label{Sec:conclusion}

A plethora of theoretical models have been worked out  in an effort to present a satisfactory explanation to the phenomenon of cosmic acceleration. Out of these models a substantial part belongs to the so called scalar-tensor theories, which contain an additional scalar DoF. In light of the upcoming observational data, the  EFT formalism is a  promising avenue as it offers a unified and model independent way to study the dynamics of linear perturbations in a wide range of scalar tensor theories possessing a  well defined Jordan frame.

Until now, when considering the EFT in the unitary gauge, the curvature perturbation $\zeta$ has been the main focus of investigation when considering the question of stability. However, this choice of variable is gauge dependent, hence one might question if going to  a gauge independent one the viable parameter space of the model changes and, most importantly, if such a gauge invariant quantity can be defined as the one describing the dynamical dark energy field. This motivated us to look for and construct a  gauge independent quantity and, consequently, to perform a comparison with the results for the original field, $\zeta$.

In this paper,  we first proceeded to define a gauge invariant quantity which describes the linear density perturbation of the dark energy field. Such a definition is very general and applicable both in the  presence of matter fields and in the late time universe. Then, moving  to the explicit stability study of the scalar DoF, we focused on avoiding the usual set of instabilities namely ghost, gradient and tachyonic instabilities for both scalar and tensor modes. These are  related to the sign of the kinetic term, the speed of propagation at high-k and the mass term at low-k respectively. Additionally, we studied the effect  of  sub-leading term in the high-$k$ expansion  as it might become important when the speed of propagation is small. Dubbed the effective undamped mass it can become problematic when it is strongly negative as the corresponding modes are unstable. Moreover, we showed that,  by doing a field redefinition in the second order action from the curvature perturbation $\zeta$  to the dark energy field, the constrains arising by imposing the absence of ghost and gradient instabilities do not change. On the contrary, the mass terms are distinctively different. Hence, in order to set the proper condition for the avoidance of tachyonic instability one needs to consider the mass term  of the dark energy field which has a real physical interpretation. In order to have an idea of the behaviour of the mass term, we studied modifications of gravity on a dS background and then we set and discuss the proper conditions one has to impose in order to ensure a stable dS. The existence of stable dS solutions is of value as it is expected to be the very late time stage of the universe. As we wished to  achieve model independent results we employed the aforementioned EFT  while neglecting any matter components due to their heavily sub-leading behaviour. 

The general,  all-encompassing nature of the original EFT action dictates that a unique approach is not feasible as sub-cases might show up which need to be treated separately. In other words, it is not possible to construct the conditions for all operators present and then reduce them to  sub-cases  by simply setting EFT functions (or combinations of them) to zero. 
This is  due to the higher spatial derivative operators. Instead, it is necessary to consider a number of sub-cases separately. We identified three main cases that deserved our attention: the case with all operators active, the beyond Horndeski class of models and the case encompassing  low-energy Ho\v rava gravity.

 We proceeded to study the stability of these three sub-cases by deriving  the kinetic term and the speed of propagation. By demanding them to be possible one guarantees the theory is free of ghosts and the gradient instability.  Additionally, we supplemented them with the same conditions guaranteeing a stable tensor sector. As discussed we find that the parameter space
identified by the no-ghost and gradient conditions is independent of the field chosen to describe the scalar DoF. In the general case when considering the low $k/a$ limit it becomes clear that  the two fields satisfy a different equation of motion. The curvature perturbations is conserved at those scales as the equation does not contain any mass term. On the contrary, the equation for the  gauge invariant dark energy field appears to have a mass term which is positive and of the same order of $H_0^2$, hence a tachyonic instability does not develop and the solution is an exponentially decaying mode. We can infer the same conclusion from the analysis of the beyond Horndeski sub-case. On the other hand,  we find that for the Ho\v rava like class  both the curvature perturbation and the gauge invariant dark energy field satisfy the same equation of motion with a  mass term dependent on the theory. Thus we have to require that $|\bar{m}^2|\leq H_0^2$ in order to guarantee a stable dS universe.

In the high-$k$ limit usually only the leading order is considered, identified as the speed of propagation. Constraining this to be positive is usually considered to be enough to guarantee stability of the corresponding modes. We proceeded to expand this analysis by not neglecting the next to leading order contribution, a term we dubbed the effective undamped mass. This term turns out be relevant for theories with a very small speed of propagation as it can  become the source of an instability. Thus, in such a case, one needs to impose an additional constraint.

As a final comment we would like to emphasize that the speed of propagation  was never identically zero . This is an interesting results  when considering the Horndeski class of models as  it was claimed that they do not propagate a scalar DoF in dS \cite{Linder:2015rcz}. While this can happen for specific cases, such as the Cubic Galileon and any analytic K-essence model, the statement does not hold in its full generality. To name one, the very well known Covariant Galileon theory has been studied and shown to propagate a DoF. To complete its study we  presented a parameter choice which not only propagates a DoF but also guarantees a stable dS background. 

Let us stress that we believe our results on the mass can be applicable  not only to a final dS stage but also at present time ($z \sim 0$), as the dark energy field is found to be dominating. Thus ensuring that the dark energy field is not developing a tachyonic instability. However, in order to guarantee that this field remains stable along all the whole evolution of the universe, one has to properly derive the mass coefficient when the matter fluids are considered. A first attempt in this direction has been done in ref. \cite{DeFelice:2016ucp}, where a dust fluid has been considered. In order to provide the mass associated to the perturbed dark energy density field one should construct the Hamiltonian for all the fields (perturbed dark energy density+ fluid densities), then work out the associated eigenvalues and finally apply the analysis in ref. \cite{DeFelice:2016ucp}. In light of the results of this work it might be important to investigate also the behaviour of an effective undamped mass term at high-$k$. Finally, although, the gauge invariant quantity we have defined for the dark energy density perturbation is still valid in the presence of matter fields, disentangling its the dynamics from the one of the others fields is an hard task and deserves a further investigation.

\begin{acknowledgments}

	ADF is grateful
  to the department of physics of Lisbon University, where this work
  was advanced, for warm hospitality. ADF was supported by JSPS
  KAKENHI Grant Numbers 16K05348, 16H01099. The research of NF is
  supported by Funda\c{c}$\tilde{\textit{a}}$o para a
  Ci$\hat{\textit{e}}$ncia e a Tecnologia (FCT) through national funds
  (UID/FIS/04434/2013) and by FEDER through COMPETE2020
  (POCI-01-0145-FEDER-007672).  GP acknowledges support from the D-ITP
  consortium, a program of the Netherlands Organisation for Scientific
  Research (NWO) that is funded by the Dutch Ministry of Education,
  Culture and Science (OCW). NF and GP acknowledge the COST Action
  (CANTATA/CA15117), supported by COST (European Cooperation in
  Science and Technology).
\end{acknowledgments}

\appendix
\section{Notation}\label{App:list}
In this Appendix we will explicitly list all the coefficients used in the main text. 

The kinetic term in action (\ref{action1}) reads
\be
\mathcal{L}_{\dot{\zeta}\dot{\zeta}}(t,k)= \f{\mathcal{A}_1(t)+\f{k^2}{a^2}\mathcal{A}_4(t)}{\mathcal{A}_2(t)+\f{k^2}{a^2}\mathcal{A}_3(t)}\,,
\ee
where
\ba
&&\mathcal{A}_1(t)= \left(F_1-3 F_4\right) \left(3 F_2^2+F_1 F_3\right)\,,\nn\\
&&\mathcal{A}_2(t)= 2\left(F_2^2+F_3 F_4\right)\,, \nn\\
&&\mathcal{A}_3(t)= 16 F_4 m_{2}^2\,,\nn\\
&&\mathcal{A}_4(t)=8 F_1 m_{2}^2\left(F_1-3 F_4\right)\,, 
\ea
and the gradient term is
\ba
G(t,k)= \f{\mathcal{G}_1(t)+\f{k^2}{a^2}\mathcal{G}_2(t)+\f{k^4}{a^4}\mathcal{G}_3}{(\mathcal{A}_2(t)+\f{k^2}{a^2}\mathcal{A}_3(t))^2}\,,
\ea
where
\ba
&&\mathcal{G}_1(t)= 4\Big[F_2 \left(F_2^2+F_3 F_4\right) \left(F_1-3 F_4\right) H \left(2 \hat{M}^2+m_0^2 (\Omega +1)\right)+2 \left(F_2{}^3 \left(\left(\dot{F}_1-3 \dot{F}_4\right) \hat{M}^2+\left(F_1-3 F_4\right) 2\hat{M}\dot{\hat{M}}\right)\r.\nn\\
&&+\l.F_2 \left(F_4 \left(3 F_4-F_1\right) \dot{F}_3 \hat{M}^2+F_3 \left(-6 F_4{}^2 \hat{M}\dot{\hat{M}}+F_4 \left(\dot{F}_1 \hat{M}^2+F_1 2\hat{M}\dot{\hat{M}}\right)-F_1 \dot{F}_4 \hat{M}^2\right)\right)-F_2^2 \left(F_1-3 F_4\right) \dot{F}_2 \hat{M}^2\r.\nn\\
&&\l.+F_3 F_4 \left(F_1-3 F_4\right) \dot{F}_2 \hat{M}^2\right)+m_0^2 \left(-\left(-F_2{}^3 \left(F_1 \dot{\Omega}-3 F_4 \dot{\Omega}+(\Omega +1) \left(\dot{F}_1-3 \dot{F}_4\right)\right)+F_2 \left(F_4 \left(F_3 \left(3 F_4 \dot{\Omega}-(\Omega +1) \dot{F}_1\right)\r.\r.\r.\r.\nn\\
&&\l.\l.\l.\l.-3 F_4 (\Omega +1) \dot{F}_3\right)+F_1 \left(F_3 \left((\Omega +1) \dot{F}_4-F_4 \dot{\Omega}\right)+F_4 (\Omega +1) \dot{F}_3\right)\right)+F_2{}^2 (\Omega +1) \left(\left(F_1-3 F_4\right) \dot{F}_2+2 F_3 F_4\right)\r.\r.\nn\\
&&\l.\l.+F_3 F_4 (\Omega +1) \left(F_3 F_4-\left(F_1-3 F_4\right) \dot{F}_2\right)+F_2{}^4 (\Omega +1)\right)\right)\Big] \,,\nn\\
&&\mathcal{G}_2(t)=8 \left(4 m_0^2 \left(-F_4 (\Omega+1) \left(F_3 F_4 \left(2 m_2^2-\hat{M}^2\right)-\left(F_1-3 F_4\right) m_2^2 \dot{F}_2\right)+F_4 F_2^2 (-(\Omega +1)) \left(2 m_2^2-\hat{M}^2\right)\r.\r.\nn\\
&&\l.\l.+3 F_4 \left(F_1-3 F_4\right) F_2 H m_2^2 (\Omega +1)+F_2 \left(F_4 \left(3 F_4 \left((\Omega +1)2 m_2\dot{m}_2-m_2^2 \dot{\Omega}\right)+m_2^2 (\Omega +1) \dot{F}_1\right)+F_1 \left(F_4 \left(m_2^2 \dot{\Omega}\r.\r.\r.\r.\r.\nn\\
&&\l.\l.\l.\l.\l.-(\Omega +1) 2m_2\dot{m}_2\right)-m_2^2 (\Omega +1) \dot{F}_4\right)\right)\right)+4 \left(6 F_2 F_4 \left(F_1-3 F_4\right) H m_2^2 \hat{M}^2+F_4 \hat{M}^2 \left(F_3 F_4 \hat{M}^2+2 \left(F_1-3 F_4\right) m_2^2 \dot{F}_2\right)\r.\r.\nn\\
&&\l.\l.-2 F_2 \left(F_4^2 \left(6 m_2^2 \hat{M}\dot{\hat{M}}-3 \hat{M}^2 m_2^2\right)-F_4 \left(m_2^2 \left(\dot{F}_1 \hat{M}^2+2F_1 \hat{M}\dot{\hat{M}}\right)-2F_1 \hat{M} m_2\dot{m}_2\right)+F_1 m_2^2 \dot{F}_4 \hat{M}^2\right)+F_2^2 F_4 \hat{M}^4\right)\r.\nn\\
&&\l.+m_0^4 F_4 \left(F_2^2+F_3 F_4\right) (\Omega +1)^2\right) \,,\nn\\
&&\mathcal{G}_3(t)=64 F_4^2 m_2^2 \left(-4 m_0^2 (\Omega +1) \left(m_2^2-\hat{M}^2\right)+4 \hat{M}^4+m_0^4 (\Omega+1)^2\right)
\ea
The kinetic and Gradient coefficients here are in a FLRW universe.

Here, we define the coefficients of action (\ref{actiondeltaphi}) 
\begin{equation}
S=\int d^4x\,a^3 \left[ \frac{a^2}{k^2}\left(Q\,{\dot\delta}_\phi^2-\mathcal{G}\,\frac{k^2}{a^2}\,\delta_\phi^2\right)\right] ,
\end{equation}
with 
\begin{eqnarray}
Q&\equiv&\frac {\Lzz^2 \left(G{\alpha_{{3}}}^{2}\frac{{k}^{2}}{a^2}-  [ H ( \eta_{\mathcal{L}}-\eta_{{3}}+\eta_{{6}}+3 ) \alpha_{{3}}-\alpha_{{6}} ] \alpha_{{6}}\Lzz
 \right) \frac{k^2}{a^2}}{ \left( \alpha_{{6}} \left( H \left( \eta_{{3}}-\eta_{{6}}-\eta_{\mathcal{L}}-3 \right) 
\alpha_{{3}}+\alpha_{{6}} \right) \Lzz+\frac{{k}^{2}}{a^2}G{\alpha_{{3}}}^{2} \right) ^{2}}\,,\\
\mathcal{G}&\equiv&\frac{\Lzz}{\left[ \alpha_{{6}} \left( H \left( \eta_{{3}}-\eta_{{6}}-\eta_{\mathcal{L}}-3 \right) \alpha_{{3}}+\alpha_{{6}}\right) \Lzz+\frac{{k}^{2}}{a^2}G{\alpha_{{3}}^2} \right] ^{\!2}}
\left( G^{2}\alpha_{{3}}^{2}\frac{{k}^{4}}{a^4}
+G\, \{  [ {\eta_{\mathcal{L}}}^{2}+(5 -2\,\eta_{{3}}-\eta_{{G}}+s_{\mathcal{L}}) \eta_{\mathcal{L}}+{\eta_{{3}}^2}\right.\nonumber\\
&&{}+
(\eta_{{G}}-s_{{3}}-5) \eta_{{3}}-3\,\eta_{{G}}+6 ] {H}^{2}{\alpha_{{3}}}^{2}
+3\,H ( \eta_{{3}}-\eta_{{6}}+1/3\,\eta_{{G}}-2/3\,\eta_{\mathcal{L}}-5/3) \alpha_{{6}}\alpha_{{3}}
+{\alpha_{{6}}}^{2} \} \frac{{k}^{2}}{a^2}\Lzz\nonumber\\
&&{}+{H}^{2}\eta_{{6}}
 [ H({\eta_{{3}}}^{2}\alpha_{{3}}-\eta_{{3}}\eta_{{6}}\alpha_{{3}}
-2\,\eta_{\mathcal{L}}\eta_{{3}}\alpha_{{3}}+\eta_{{3}}\alpha_{{3}}s_{{3}}-
\eta_{{3}}\alpha_{{3}}s_{{6}}+\eta_{\mathcal{L}}\eta_{{6}}\alpha_{{3}}+{\eta_{\mathcal{L}}}^{2}\alpha_{{3}}
+\eta_{\mathcal{L}}\alpha_{{3}}s_{{6}}\nonumber\\
&&{}-\left. \eta_{\mathcal{L}}\alpha_{{3}}s_{\mathcal{L}}-6\,\eta_{{3}}\alpha_{{3}}+3\,\alpha_{{3}}\eta_{{6}}
+6\,\alpha_{{3}}\eta_{\mathcal{L}}+3\,\alpha_{{3}}s_{{6}}+9\,\alpha_{{3}})+\alpha_{{6}}\eta_{{6}}-\alpha_{{6}}\eta_{\mathcal{L}}-\alpha_{{6}}s_{{6}}-3
\,\alpha_{{6}} ] \alpha_{{6}}{\Lzz^2} \right) ,
\end{eqnarray}
and %
\begin{equation}
s_{\mathcal L}\equiv\frac{\dot\eta_{\mathcal{L}}}{H\eta_{\mathcal{L}}}\,,\quad s_3\equiv\frac{\dot\eta_3}{H\eta_3}\,,\quad s_6\equiv\frac{\dot\eta_6}{H\eta_6}\,,\quad \eta_G=\frac{\dot G}{H\,G}\,.
\end{equation}

Moreover, the explicit expressions for the $\alpha_i$ and $\mu_i$ coefficients in the dS limit used in Sec. \ref{Sec:dS}  are
\ba\label{alphafull}
&&\alpha_3(t,k)= -\frac{2 \left(F_1-3 F_4\right) \left(2 F_2 H_0+F_3+8 \f{k^2}{a^2} m_{2}^2\right)}{3 H_0 \left(F_2^2+F_3 F_4+8 F_4 \f{k^2}{a^2} m_{2}^2\right)}\nn \\
&&\alpha_6(t,k)= \f{2k^2}{a(t)^2}\frac{ H_0\l( F_4 H_0 -2 F_2  \r)\left(2 \hat{M}^2+m_0^2 (\Omega +1)\right)+F_2^2+F_3 F_4+8 F_4 \f{k^2}{a^2} m_{2}^2}{3 H_0^2  \left(F_2^2+F_3 F_4 +8 F_4 \f{k^2}{a^2} m_{2}^2\right)}\,,
\ea
and
\ba\label{coefficientphimu3full}
&&\mu_3(t,k)=\l\{\mathcal{L}_{\dot{\zeta}\dot{\zeta}} \left(-6 H_0 \alpha _6 \alpha _3 \dot{\mathcal{L}}_{\dot{\zeta}\dot{\zeta}}+\alpha _6 \alpha _3 \ddot{\mathcal{L}}_{\dot{\zeta}\dot{\zeta}}+\alpha _6{}^2 \dot{\mathcal{L}}_{\dot{\zeta}\dot{\zeta}}+2 \alpha _6 \dot{\mathcal{L}}_{\dot{\zeta}\dot{\zeta}} \dot{\alpha}_3\right)+\mathcal{L}_{\dot{\zeta}\dot{\zeta}}^2 \left(3 H_0 \alpha _6 \left(2 \dot{\alpha}_3+\alpha _6\right)-9 H_0^2 \alpha _3 \alpha _6 \r.\r. \nn\\
&&\l.\l.-\alpha _6 \left(2 \dot{\alpha}_6+\ddot{\alpha}_3\right)+\alpha _3 \ddot{\alpha}_6\right)+2 \alpha _3 \dot{\mathcal{L}}_{\dot{\zeta}\dot{\zeta}} \left(-\alpha _6 \dot{\mathcal{L}}_{\dot{\zeta}\dot{\zeta}}\right)-\f{k^2}{a^2} \mathcal{L}_{\dot{\zeta}\dot{\zeta}} \alpha _3{}^2 \dot{G} +\f{k^2}{a^2} G \alpha _3 \left(\mathcal{L}_{\dot{\zeta}\dot{\zeta}} \left(5 H_0 \alpha _3-2 \dot{\alpha}_3\right)+2 \alpha _3 \dot{\mathcal{L}}_{\dot{\zeta}\dot{\zeta}}\right)\r\}\nn\\
&&\times 1/\l\{\mathcal{L}_{\dot{\zeta}\dot{\zeta}} \left( \mathcal{L}_{\dot{\zeta}\dot{\zeta}} \left(\alpha _6{}^2-3 H_0 \alpha _3 \alpha _6+\alpha _6 \dot{\alpha}_3-\alpha _3 \dot{\alpha}_6\right)-\alpha _6 \alpha _3 \dot{\mathcal{L}}_{\dot{\zeta}\dot{\zeta}}+\f{k^2}{a^2} G \alpha _3{}^2\right)\r\}\,,
\ea
\ba\label{coefficientphimu6full}
&&\mu_6(t,k)=\l\{a^2 \left(a^2 \left(\mathcal{L}_{\dot{\zeta}\dot{\zeta}} \left(-3 H_0 \alpha _3 \left(-2 \dot{\mathcal{L}}_{\dot{\zeta}\dot{\zeta}} \dot{\alpha}_6\right)-\dot{\mathcal{L}}_{\dot{\zeta}\dot{\zeta}} \left(2 \dot{\alpha}_3+\alpha _6\right) \dot{\alpha}_6+\alpha _3 \left(-\ddot{\mathcal{L}}_{\dot{\zeta}\dot{\zeta}} \dot{\alpha}_6+\dot{\mathcal{L}}_{\dot{\zeta}\dot{\zeta}} \ddot{\alpha}_6\right)\right)\r.\r.\r.\nn\\
&&\l.\l.\l.+\mathcal{L}_{\dot{\zeta}\dot{\zeta}}^2 \left(-3 H_0 \left(\alpha _6 \dot{\alpha}_6+2 \dot{\alpha}_3 \dot{\alpha}_6-\alpha _3 \ddot{\alpha}_6\right)+9 H_0^2 \alpha _3 \dot{\alpha}_6+\dot{\alpha}_6 \ddot{\alpha}_3-\left(\dot{\alpha}_3+\alpha _6\right) \ddot{\alpha}_6+2 \dot{\alpha}_6{}^2\right)
+\alpha _3 \dot{\mathcal{L}}_{\dot{\zeta}\dot{\zeta}} \left(2 \dot{\mathcal{L}}_{\dot{\zeta}\dot{\zeta}} \dot{\alpha}_6\right)\right)\r.\r.\nn\\
&&\l.\l.+k^2 \alpha _3 \dot{G} \left(\mathcal{L}_{\dot{\zeta}\dot{\zeta}} \left(-3 H_0 \alpha _3+\dot{\alpha}_3+\alpha _6\right)-\alpha _3 \dot{\mathcal{L}}_{\dot{\zeta}\dot{\zeta}}\right)\right)+k^2 a^2 G \left(\alpha _3 \left(5 H_0 \alpha _3 \dot{\mathcal{L}}_{\dot{\zeta}\dot{\zeta}}+\alpha _3 \ddot{\mathcal{L}}_{\dot{\zeta}\dot{\zeta}}-2 \alpha _6 \dot{\mathcal{L}}_{\dot{\zeta}\dot{\zeta}}-2 \dot{\mathcal{L}}_{\dot{\zeta}\dot{\zeta}} \dot{\alpha}_3\right)\r.\r.\nn\\
&&\l.\l.+\mathcal{L}_{\dot{\zeta}\dot{\zeta}} \left(-5 H_0 \alpha _3 \left(\dot{\alpha}_3+\alpha _6\right)+6 H_0^2 \alpha _3{}^2-3 \alpha _3 \dot{\alpha}_6+2 \dot{\alpha}_3{}^2+3 \alpha _6 \dot{\alpha}_3-\alpha _3 \ddot{\alpha}_3+\alpha _6{}^2\right)\right)+k^4 G^2 \alpha _3{}^2\r\}\times \nn\\
&&1/\l\{a^2 \mathcal{L}_{\dot{\zeta}\dot{\zeta}} \left(a^2 \left(\mathcal{L}_{\dot{\zeta}\dot{\zeta}} \left(-3 H_0 \alpha _3 \alpha _6+\alpha _6 \dot{\alpha}_3-\alpha _3 \dot{\alpha}_6+\alpha _6{}^2\right)-\alpha _6 \alpha _3 \dot{\mathcal{L}}_{\dot{\zeta}\dot{\zeta}}\right)+k^2 G \alpha _3{}^2\right)\r\} \,.
\ea

\end{document}